\documentclass[preprint,review,12pt]{elsarticle}

\usepackage{graphics}
\usepackage{graphicx}

\usepackage{amssymb}

\usepackage{lineno}

\usepackage{amsmath}
\usepackage[version=4]{mhchem}

\usepackage{listings}
\usepackage{color}

\usepackage{booktabs}
\usepackage{float}

\definecolor{dkgreen}{rgb}{0,0.6,0}
\definecolor{gray}{rgb}{0.5,0.5,0.5}
\definecolor{mauve}{rgb}{0.58,0,0.82}

\newcommand{\ufive}{$^{235}$U}
\newcommand{\ueight}{$^{238}$U}
\newcommand{\punine}{$^{239}$Pu}
\newcommand{\puone}{$^{241}$Pu}

\newcounter{bla}

\begin{document}

\begin{frontmatter}



\title{CONFLUX: A Standardized Framework to Calculate Reactor Antineutrino Flux}


\author[a]{Xianyi Zhang\corref{author}}
\author[b]{Anosh Irani}
\author[a]{Michael P. Mendenhall}
\author[b]{Nathan Rybicki}
\author[c]{Leendert Hayen}
\author[a]{Nathaniel Bowden}
\author[d]{Patrick Huber}
\author[b]{Bryce Littlejohn}
\author[e]{Sandra Bogetic}

\address[a]{Lawrence Livermore National Laboratory, Livermore, CA, USA}
\address[b]{Illinois Institute of Technology, Chicago, IL, USA}
\address[c]{The laboratory of corpuscular physics of Caen, Caen, France}
\address[d]{Virginia Polytechnic Institute and State University, Blacksburg, VA, USA}
\address[e]{University of Tennessee, Knoxville, TN, USA}

\cortext[author]{Corresponding author. \textit{E-mail address:} zhang39@llnl.gov}

\begin{abstract}
Nuclear fission reactors are abundant sources of antineutrinos for neutrino physics experiments.  
The flux and spectrum of antineutrinos emitted by a reactor can indicate its activity and composition, suggesting potential applications of neutrino measurements beyond fundamental scientific studies that may be valuable to society.  
The utility of reactor antineutrinos for applications and fundamental science is dependent on the availability of precise predictions of these emissions.  
For example, in the last decade, disagreements between reactor antineutrino measurements and models have inspired revision of reactor antineutrino calculations and standard nuclear databases as well as searches for new fundamental particles not predicted by the Standard Model of particle physics.  
Past predictions and descriptions of the methods used to generate them are documented to varying degrees in the literature, with different modeling teams incorporating a range of methods, input data, and assumptions.  
The resulting difficulty in accessing or reproducing past models and reconciling results from differing approaches complicates the future study and application of reactor antineutrinos.    
The CONFLUX (Calculation Of Neutrino FLUX) software framework is a neutrino prediction tool built with the goal of simplifying, standardizing, and democratizing the process of reactor antineutrino flux calculations.
CONFLUX include three primary methods for calculating the antineutrino emissions of nuclear reactors or individual beta decays that incorporate common nuclear data and beta decay theory.  
The software is prepackaged with the current nuclear databases, including ENDF.B/VIII, JEFF-3.3, and ENSDF, and it includes the capability to predict time-dependent reactor emissions, adjust nuclear database or beta decay inputs/assumptions, and propagate related sources of uncertainty.  
This paper describes the CONFLUX software structure, details the methods used for flux and spectrum calculations, and provides examples of potential use cases. 
\end{abstract}

\end{frontmatter}



{\bf PROGRAM SUMMARY}

\begin{small}
\noindent
{\em Program Title: CONFLUX}                                          \\
{\em CPC Library link to program files:} (to be added by Technical Editor) \\
{\em Developer's repository link:} https://github.com/CNFLUX/conflux \\
{\em Code Ocean capsule:} (to be added by Technical Editor)\\
{\em Licensing provisions:} MIT  \\
{\em Programming language: Python, C++}                                   \\
{\em Supplementary material:}                                 \\

\end{small}

\section{Introduction}
\label{sec:Introduction}

Nuclear fission reactors are abundant sources of electron antineutrinos (henceforth simplified to ``reactor neutrinos''). 
Reactor neutrino measurements have been an essential tool for understanding fundamental aspects of particle physics~\cite{Akindele:2024nzu}, including the neutrino mass and flavor transformation properties~\cite{KamLAND_rate,KamLAND_shape,DoubleChooz:2011ymz,DayaBay:2012fng,RENO:2012mkc}, searching for particles and forces beyond the Standard Model~\cite{bib:neos,danss_osc,stereo_2018,prospect_osc}, and in the near-future the neutrino mass hierarchy~\cite{JUNO:2021vlw}.
Neutrinos also represent the only completely unattenuated form of radiation emitted by a nuclear reactor, making them a unique source of detailed information about the core's state and content that can be leveraged to perform remote evaluations~\cite{Bernstein:2019hix,Carr_Review}.  
Both scientific measurements and monitoring applications of neutrinos require an accurate prediction of reactor neutrino flux and spectrum.
For example, certain types of physics beyond the Standard Model can be probed with unprecedented precision at reactors via comparison of measured and predicted rates of neutrino-nucleus coherent elastic scattering inside reactor-adjacent detectors~\cite{ricochet,connie,conus} -- a comparison directly dependent on accurate modeling of neutrino production rates in the core.  
In the neutrino applications space, potential cases such as remote reactor evaluation rely on accurate reference neutrino source terms to determine, for example, the contributors of fission in the monitored core~\cite{NK_reactor,Christensen:2014pva,Erickson_reactor}.

Currently, significant discrepancies ranging from the few percent to the 10s of percent level or higher, are observed between calculated reactor neutrino fluxes and spectra and direct reactor neutrino measurements. 
Measured reactor neutrino fluxes are a few percent low compared to data driven flux models~\cite{mueller_improved_2011, mention_reactor_2011, huber_determination_2011}.  
In addition, spectrum shape differences have been observed between the data of multiple experiments and predictions: these deviations range from the 10\% level around 5-7~MeV neutrino energy~\cite{bib:prl_reactor,bib:reno_shape,bib:neos} to more than 100\% at the highest reactor neutrino energies~\cite{DayaBay:2022eyy}.  
Despite the gradually improving and diversifying state of reactor flux calculations~\cite{PhysRevLett.123.022502, PhysRevC.100.054323, PhysRevD.104.L071301, giunti_reactor_2022, Letourneau:2022kfs, Perisse:2023efm} and the improving granularity and breadth of acquired direct reactor neutrino measurements including new measurements of individual isotopic contributions from \ce{^235U} and \ce{^239Pu}~\cite{DayaBay:2017jkb, DayaBay:2019yxq, DayaBay:2021dqj, prospect_collaboration_improved_2021, almazan_stereo_2023} and even \ce{^238U}~\cite{surukuchi_flux,Fujikake:2023luo}, the community has yet to fully resolve the source of these discrepancies.  

The major fissile isotopes, \ce{^235U}, \ce{^238U}, \ce{^239Pu}, and \ce{^241Pu}, generate $\sim1000$ fission products and more than 500 of these are beta-unstable.  
The neutrino emissions of these beta-unstable fission products -- the source of most reactor-emitted neutrinos -- have been modeled through two common methods: the `summation' approach (also referred to as the `ab initio' approach) builds predicting neutrino spectra directly from known or assumed properties of all beta-unstable fission products, while the `conversion' approach generates neutrino spectra using aggregate fission beta measurements as a primary input.  
Non-fissile beta-decaying actinides and lower mass activated beta-decaying nuclides also contribute non-negligibly to reactor neutrino production~\cite{PROSPECT:2020vcl}; treatment of which requires additional reactor-specific modeling and theoretical calculation.

A summation calculation of reactor neutrino emissions relies on a vast array of nuclear data, such as fission product yields (FPY) of fissile isotopes and nuclear structure attributes (nuclear level energies and spin-parities and beta-decay branching fractions) for the various fission products.
The neutrino spectrum from the decay branches of each fission product nuclide needs to be calculated with theoretical corrections specific to those nuclear transitions.  
The summation method can be used to account for neutrino contributions from all decay branches of all fission products for any fissile isotope, as well as contributions from beta-decaying non-fissile isotopes in a reactor.  
This method is accompanied by an array of challenges, uncertainties, and limitations, including missing beta-decay information, uncertain factors for theoretical corrections, uncertainties in FPYs and beta feedings, and ill-defined FPY uncertainty correlations.  
This complex theoretical approach, used in published calculations~\cite{PhysRevD.104.L071301, PhysRevLett.105.181801, PhysRevLett.123.022502}, will be detailed further in Section~\ref{sec:BSG}.  

The alternate `conversion' neutrino flux prediction method makes use of existing aggregate fission beta measurements for the key fissile isotopes  \ce{^235U}, \ce{^239Pu}, and \ce{^241Pu}~\cite{von_feilitzsch_experimental_1982, hahn_antineutrino_1989}.  
These predictions convert measured beta energy spectra into predicted neutrino spectra using basic energy conservation and other nuclear theory principles~\cite{huber_determination_2011}.
Because the contributions of individual beta branches in the aggregate  beta spectrum cannot be known, conversion predictions are performed by fitting a limited number of virtual beta spectra with variable contribution rates onto the aggregate spectra.  
By eschewing complete modelling of all branch's individual contributions, the conversion approach avoids nuclear data uncertainties and biases while taking on uncertainties related to the underlying beta measurements -- uncertainties that, in principle, are more easily quantifiable and less prone to bias.  
Unfortunately, recent re-measurements of \ce{^235U} and \ce{^239Pu} beta spectra suggest that biases accompanying long-used aggregate beta data may actually be quite substantial~\cite{PhysRevC.100.054323}.  
In addition, as in the summation method, the beta decay theory incorporated when modeling virtual beta branch contributions requires a range of corrections, many of which remain under active investigation.

With the advent of higher precision, more diverse global measurements of reactor antineutrinos, a third family of reactor neutrino flux calculation methods has emerged based on direct `benchmarking' measurements of aggregate neutrino spectra from large,  high-statistics inverse beta decay (IBD) experiments~\cite{bib:neos, DayaBay:2021dqj}.  
Due to the unattenuated nature of neutrino energies and emissions, this represents a  promising method for generating reliable, minimally biased neutrino flux predictions for future reactor applications and fundamental physics studies.  
However, the limited statistics, neutrino energy resolution, and range of sampled reactor types of existing IBD-based reactor neutrino measurements place a  limitation on the near-term applicability of this method.

The present landscape of reactor neutrino flux predictions are non-standardized and hard to access.  
Existing reactor neutrino calculations in the literature performed using the three methods described above are authored by a range of different groups with varying degrees of accompanying description.  
Calculations are most often shared as individual data products or files not accompanied by the tools or code used to generate them.  
For novice of nuclear physics or nuclear databases, this state of affairs results in reactor neutrino flux predictions being treated as a ``black box'', with most turning to a small number of canonical summation or conversion predictions~\cite{mueller_improved_2011,huber_determination_2011,PhysRevLett.123.022502}.  
For modeling experts, this state of affairs makes neutrino flux calculations difficult to repeat due to differing ways to process the nuclear data and assumed input variables.  
Moreover, methods and nuclear data are updated frequently, adding further complexity to efforts to compare one calculation to another.

Looking to the future, it seems likely that re-evaluations of both the underlying nuclear theory and the data used in corresponding calculations will be necessary to understand the origins of this persistent problem.   
Future calculations of reactor neutrino flux will rely on frequent crosschecks with updated nuclear database tabulations, updated nuclear decay data supplied by new measurements~\cite{tas_lots,algora_beta-decay_2021}, and customized  theoretical functions or treatments.  
Comparisons between different prediction methods will also be necessary to assess the strengths and weaknesses of different approaches and improve the overall accuracy of reactor flux modeling. 
Therefore, a flux calculation software that is easy to access and provides universal propagation of experimental data, functions, and nuclear databases would be particularly valuable for this community.
In addition, remote reactor neutrino surveys and other neutrino-based applications activity will require reliable tools for deriving reactor-specific neutrino flux references to evaluate reactor activities and/or content.  
Ideally, such a tool would be efficient, user-friendly, and customizable to accommodate the needs of neutrino non-experts and experts alike.

With these considerations and current limitations in mind, we have developed an accessible new code base for generating reactor antineutrino flux predictions.  
The CONFLUX (Calculation Of Neutrino FLUX) software package is built to standardize the input nuclear data, data and uncertainty processes, as well as theoretical beta decay models for reactor neutrino flux calculation for all three (summation, conversion, and data-driven) prediction methods.  
The purpose of building the CONFLUX framework is to make reactor neutrino flux calculations simple and repeatable while enabling a wide range of customizable factors.  
The software is written so users can also use default beta theory models and nuclear databases to calculate reactor neutrino flux with designated reactor burnup models, providing reference neutrino flux for measurements of specific reactors.  

This paper describes the structure, formats for data, and methods used to calculate neutrino flux of CONFLUX's summation and beta-conversion modes.
The structure the software and the block diagrams for neutrino flux calculation are described in section~\ref{sec:general}.
Details of the nuclear data usage, the summation methods, and corresponding uncertainty calculations are shown in section~\ref{sec:summation}.
Section~\ref{sec:b-conversion} describes the beta conversion procedures.
In section~\ref{sec:example}, use cases are given as examples of the summation and conversion calculations, including changing key values in the calculation, updating nuclear data with user input, extracting of reactor neutrino spectrum with conditions, and hybrid mode calculations.
The paper concludes with the potential CONFLUX contributions to the neutrino and nuclear physics fields, and future improvements.

\section{General Information}
\label{sec:general}
\begin{figure}[htbp!]
    \centering
    \includegraphics[width=0.9\textwidth]{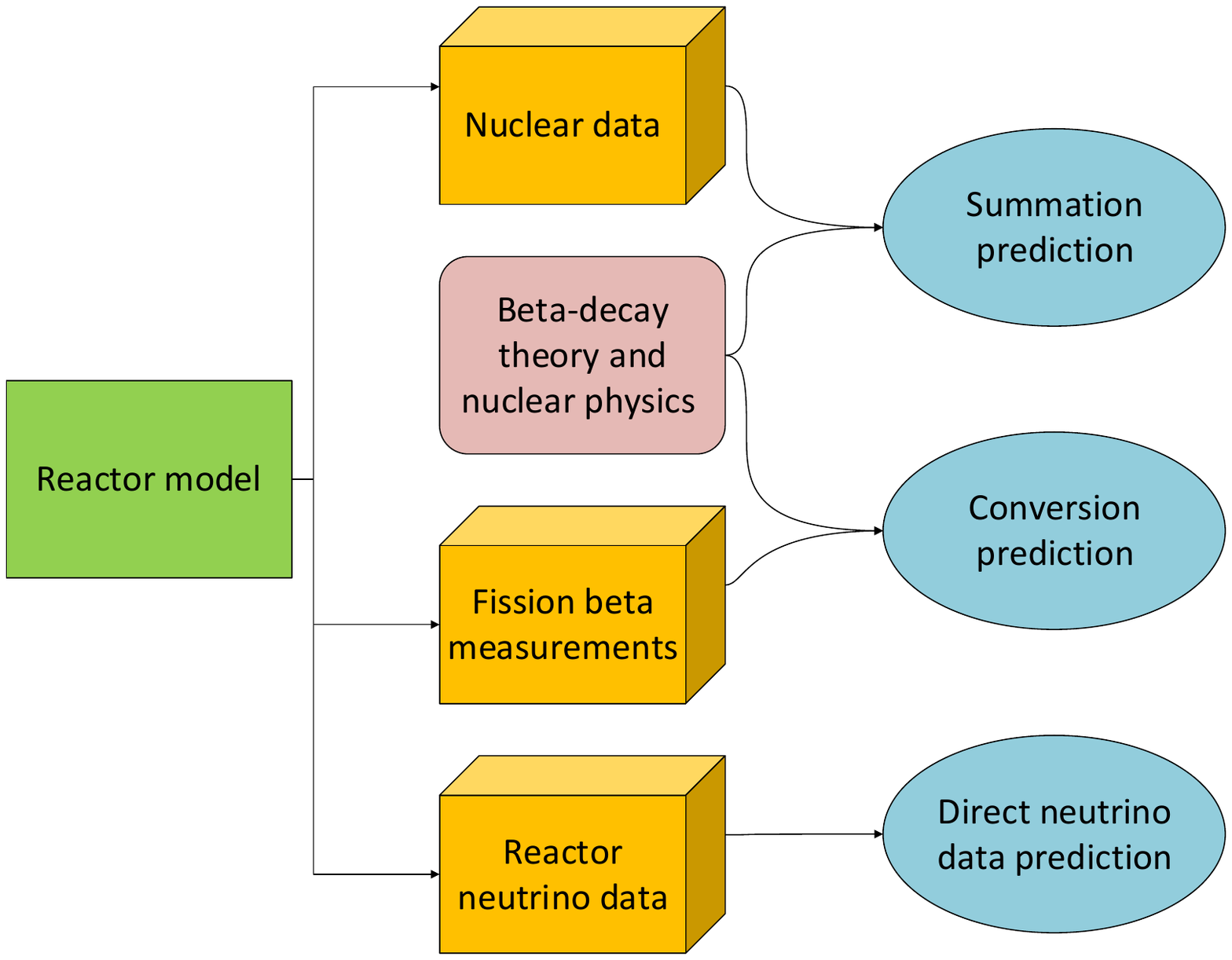}
    \caption{Basic calculation modules of CONFLUX. Users will provide a reactor model composed of fission fractions and potentially their evolution with time.  The neutrinos generated by fission-produced beta-decaying isotopes in the user-provided model will be calculated through one of three user-chosen calculation methods. Pre-packaged calculation functions and nuclear databases provided within CONFLUX will be used to calculate an output neutrino spectrum.}
    \label{fig:philosophy}
\end{figure}

CONFLUX is written in order to standardize and simplify the calculation procedures and factors for reactor neutrino predictions.
The software framework is made with modules as shown in figure~\ref{fig:philosophy}.
CONFLUX is written in Python3 and packed with data in Extensible Markup Language (xml) format and Comma-Separated Value (csv) format for readability.
The prepackaged xml and csv are processed from nuclear databases  ENDF/B-VIII~\cite{brown_endfb-viii0_2018} and JEFF-3.3~\cite{plompen_joint_2020}, and published fissile isotope beta measurements~\cite{hahn_antineutrino_1989}
The python3 executable programs that process the data are also prepackaged in the module so users can update the input nuclear databases and continue using CONFLUX. 
CONFLUX is compatible to the PyPl library, so the module can be installed for usage outside of the package for general python programs.
One major goal of CONFLUX is to enable customization of beta decay theoretical models (such as corrections for forbidden transition shape factors), incorporated nuclear data, and other reactor or fission product related assumptions, so that one can study the impacts of chosen models, input data, and assumptions in a controlled manner.  
Nuclear data, experimental beta-decay data, as well as functions for theoretical beta-decay calculation are shared among the summation and the conversion modes.

\begin{figure}[ht]
    \centering
    \includegraphics[width=0.5\textwidth]{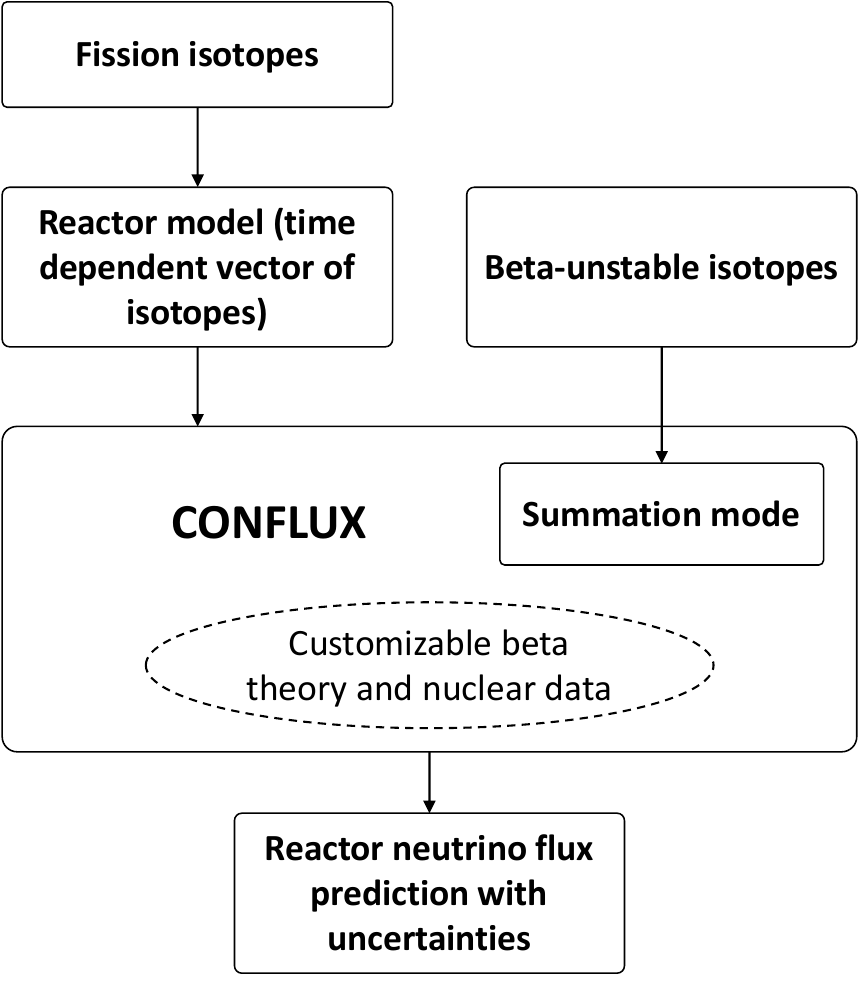}
    \caption{A high-level block diagram depicting required inputs and provided outputs of the CONFLUX software package.}
    \label{fig:procedure}
\end{figure}

The calculation procedure is illustrated in Figure~\ref{fig:procedure}.  
The procedure begins with specification of user inputs for the reactor model, including the rate of fission for the primary fission isotopes and rates of decay for other beta-unstable isotopes in the core not generated by fission.
The specified fission isotope contributions then go through the beta/neutrino spectrum calculation of the user-defined calculation mode. 
Non-fission beta-unstable isotope contributions to neutrino flux are calculated theoretically through the summation mode using nuclear data and the beta-decay theory.
The result of the calculation includes energy dependent flux of neutrino and uncertainty.  

\subsection{Calculation Modes}
At present, the published CONFLUX software contains two modes: the summation mode and the conversion mode.  
The summation mode that builds neutrino spectra of beta-decaying fission products from nuclear databases is detailed in Section~\ref{sec:summation}.  
The conversion mode relies on conservation of energy and momentum in beta decay to fit input aggregate fission beta spectra of fissile isotopes with theoretical beta spectra and convert the best-fit beta spectra to corresponding neutrino spectra.  
The beta spectrum data and methods used in this mode are described in Section~\ref{sec:b-conversion}.  
The third mode -- data-driven neutrino mode -- is under development for publication in the near future.   
This mode calculates reactor neutrino flux by fitting direct reactor neutrino measurement data simultaneously and searching for the best fit isotopic reactor neutrino flux model.  
The different calculation modes share common set of prepackaged nuclear data and beta decay theory inputs and assumptions.  

\subsection{User Inputs}
CONFLUX accepts as input a time-dependent vector of fissile isotope fission rates and uncertainties corresponding to the user's preferred reactor model.  
These time-dependent inputs may be borrowed from the existing literature -- i.e.~\cite{bib:prl_reactor,Bernstein:2016ayp} or may be the result of a separate dedicated reactor modelling exercise performed by the user.  
An analyzer can define a fissile isotope input with its atomic number $Z$, mass number $A$, and fission neutron energy $E_i$, in the form of {\texttt{Isotope(Z, A, E$_i$)}}.  
Once defined, the contribution and uncertainty of the isotope in the reactor can be placed to form a complete input reactor model.  

CONFLUX's neutrino calculations can also be initiated by providing an input list of non-fissile isotopes existing in a reactor.  
In this method, a list of beta-decaying isotopes containing, for example, specific fission products, non-fissile actinides, or activated reactor materials, as well as accompanying decay rates, are defined independently of any fission reactor model to allow complete summation calculations for detailed reactor simulations or for studies of the neutrino or beta spectrum attributes of a subset of radio-nuclides.  
A mixture of fissile isotope inputs and non-fissile beta decaying inputs is also allowed.  
CONFLUX is programmed to encourage users to change theoretical assumptions in beta calculations and/or provide new beta-decay or fission product data to study the impact of updated inputs on reactor neutrino production.    
Examples of using time dependent reactor fission models and different modifications of beta decay or nuclear data are included in CONFLUX, will be detailed in section~\ref{sec:example}.

CONFLUX also contains an executable at {\texttt{exec/quickflux.py}} to allow users to input calculation parameters through a JSON data format input file (macros).  
This executable allows a user to quickly calculate reactor neutrino flux with simple inputs without writing a python script.  
The spectrum shape, reactor fission and non-fission beta-decay contributions, calculation mode, and output file type can be edited in the JSON file to let {\texttt{quickflux.py}} calculate desired reactor neutrino output quickly for novice programmers and users.   

\subsection{Results of calculation}
The output of a CONFLUX calculation is one or more appropriately normalized reactor neutrino or beta energy spectra, depending on user inputs, as well as each generated spectrum's associated uncertainty.  
The formats of the output spectra and uncertainties are NUMPY arrays whose length is dependent on input x-axis settings.  
Depending on user-defined settings, these arrays can then be processed and saved into multiple data formats.  
During a CONFLUX calculation, users can choose to output all or only a portion of the neutrino flux generated by sub-sets of fissioning isotopes or beta-unstable isotopes with specified  properties, e.g., specific fission yield ranges, fission yield uncertainties, or beta branch end-point energies.  
The ability of CONFLUX to provide varied reactor neutrino flux outputs can help users to understand the impacts of calculation adjustments and uncertainties by specific groups of fission products.  

CONFLUX propagates uncertainties representing the best available documented knowledge provided by nuclear data and theories.  
Beta decay modelling uncertainties, fission product yield uncertainties, and uncertainty covariances among fission products are all considered in the summation mode.  
Uncertainty calculations for the summation mode are described in section~\ref{sec:summation}.  
Amplitudes of the beta conversion uncertainty will be driven by the uncertainty of the aggregate fission beta data.   
The covariance of best-fit virtual beta branches and converted neutrino spectra are calculated through MC sampling, a method described in section~\ref{sec:b-conversion}.

\section{Beta Spectrum Generator}
\label{sec:BSG}
CONFLUX preloads a beta-decay neutrino/beta spectrum generator engine based on the \texttt{BSG} package~\cite{hayen_beta_2019}. Specifically, a customized Python implementation was developed to take into account all corrections up to $\mathcal{O}(10^{-4})$ as described in Ref. \cite{hayen_beta_2019} for allowed decays while incorporating the $\beta$ spectrum shape changes for forbidden transitions. 
\subsection{Formalism}
The transition rate can generally be written as
\begin{equation}
    \frac{d\Gamma}{dW} = (1+\Delta_R^V) \frac{G_F^2W_0^5V_{ud}^2}{2\pi^3} pW(W_0-W)^2 C_{\rm shape} F_0L_0SR
\end{equation}
where $G_F = 1.166 \times 10^{-5}$ GeV$^{-2}$ is the Fermi constant, $V_{ud} = 0.97373(31)$ is the up-down element of the Cabibbo-Kobayashi-Maskawa matrix, $\Delta_R^V$ is a renormalization of the weak interaction through virtual photon effects, and $p (W)$ are the electron momentum (total energy) in units of $m_e c$ ($m_ec^2$). 
Additionally, $F_0L_0$ is the corrected Fermi function, $S$ takes into account atomic screening effects, $R$ contains additional radiative corrections and $C_{\rm shape}$ contains all nuclear structure information. 
For allowed decays ($\Delta J^\pi = (0, 1)^{+}$), the latter is dominated by the Gamow-Teller strength, so that one may write $C_{\rm shape}^{A} = |\mathcal{M}_{\rm GT}|^2(1+ \delta_A)$ where $\delta_A$ are percent-level corrections due to, e.g., weak magnetism \cite{PhysRevD.104.L071301}. 
For so-called non-unique forbidden transitions (i.e. transitions where $|\Delta J| < n$ and $\Delta \pi = (-1)^n$ for a degree of forbiddenness $n \geq 1$), several different matrix elements may dominate on a case-by-case scenario due to the underlying nuclear structure, resulting in a large variation of anticipated shape factors~\cite{hayes_systematic_2014}. 
Unique transitions, on the other hand, where $\Delta J = n$ are once more dominated by a single matrix element so that one may write
\begin{equation}
    C_{\rm shape}^{U} \propto \sum_{k=1}^{L} \lambda_k \frac{p^{2(k-1)}q^{2(L-k)}}{(2k-1)![2(L-k)+1]!}
\end{equation}
where $q = W_0-W$ is the antineutrino momentum, $L$ is the maximum angular momentum change and $\lambda_k$ is a non-trivial Coulomb function which depends on the electron energy. 
The importance of the latter was already noted \cite{hayen_beta_2019}, and is calculated explicitly using the methods described by B\"uhring 1984~\cite{buhring_screening_1984}. 

The \texttt{BSG} package~\cite{hayen_beta_2019} is loaded by default in CONFLUX to calculate the beta and neutrino spectrum shapes based on input atomic properties, end-point energies, and forbidden-ness of decay.  
By default, CONFLUX treats all forbidden transitions as first-order unique forbidden transitions in beta decay, and applies the weak magnetism correction with the factor $b_{Ac}=0.0047$ as described in \cite{hayen_beta_2019}.  
The spectrum shape generating function can be customized by the user through an alternative beta-decay function with the same input arguments and output calculation results.  

The \texttt{BSG} package in CONFLUX has been modified to allow non-zero neutrino mass to distort both beta and neutrino spectra.  
This functionality allows CONFLUX to serve the community of beta spectrum measurement experiments focused on probing the existence of non-zero neutrino masses~\cite{Smith:2016vku,Friedrich:2020nze,deGroot:2022tbi}.  
These efforts can use CONFLUX's calculated beta spectra as a reference to predict the distortion or discontinuity in beta spectrum shape caused by neutrino mass.  



\section{Summation Calculation}
\label{sec:summation}
\begin{figure}[ht]
    \centering
    \includegraphics[width=0.9\textwidth]{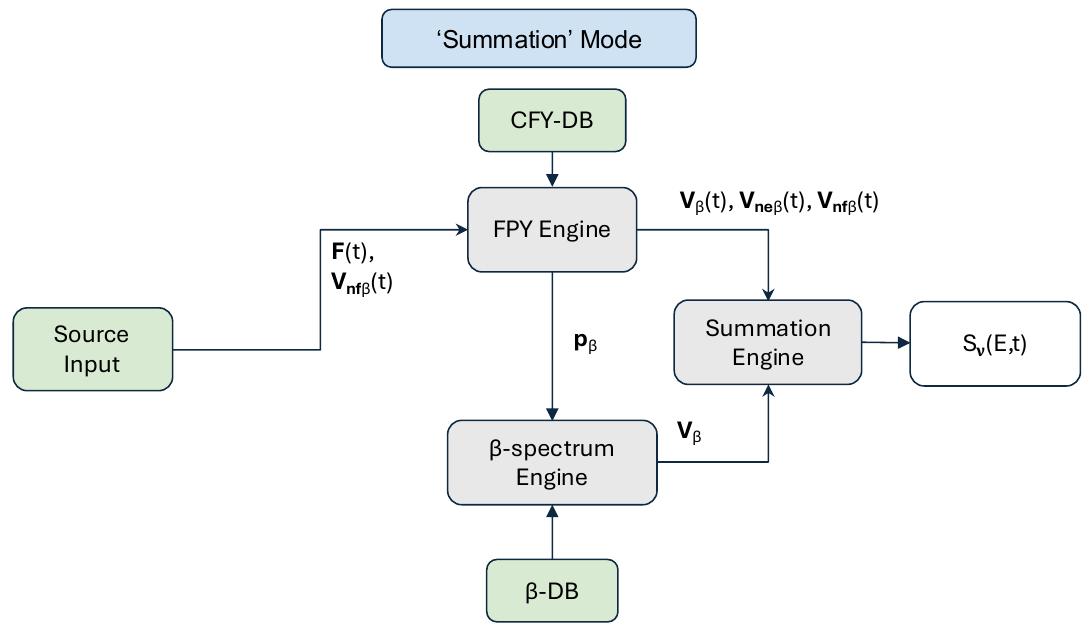}
    \caption{A block diagrams representing the processes in a summation mode calculation performed with CONFLUX.}
    \label{fig:sumblock}
\end{figure}

Fission reactors generate neutrinos through decays of beta-unstable fission products and other beta-unstable nuclides generated by non-fission processes, such as neutron activation of in-fuel actinides or non-fuel materials.  
Among fissile isotopes, \ufive, \ueight, \punine, and \puone~are the dominant sources of fission in most reactors.  
In a summation-mode calculation, CONFLUX reads parsed versions of standard nuclear fission product yield (FPY) databases, such as ENDF and JEFF, to tally the quantity of fission products generated by these isotopes.  
CONFLUX then sums together the individual beta/neutrino spectra from each branch of each fission product, as well as contributions from non-fission beta-decaying nuclides.  
Each branch's spectrum is dictated by parsed versions of standard nuclear structure databases, such as ENSDF.  
As shown in figure~\ref{fig:sumblock}, this mode involves reading in nuclear data and user-defined inputs to account for both fission and non-fission contributions, then calculating the total spectrum shape while properly normalizing by the absolute decay rate or fractional contribution each decay branch.  
Users can create contributions of all isotopes with respect to time to calculate an evolving model of neutrino production specific to their reactor.  

\subsection{Summation Processes}
The fission-produced neutrino energy spectrum generated at a specific time, $F(E, t)$, can be expressed as
\begin{equation}
\label{eq:sum}
    F(E, t) = \sum_i R_i(t)\sum_j f_{ij}S_{j}(E),
\end{equation}
where $R_i(t)$ is the time-dependent rate of fission of fissile isotope $i$ in a reactor,  $f_{ij}$ is the cumulative FPY of the produced beta-unstable isotope $j$, and $S_{j}(E)$ is the neutrino spectrum generated by $j$.
Many beta-unstable isotopes contain many beta decay branches, where the parent isotope decays into different energy levels of the corresponding daughter isotope.
Thus, the beta/neutrino spectrum of each isotope $S(E)$ can be expressed as
\begin{equation}
    S(E) = \sum_l b_l s_l(E),
\end{equation}
where $b_l$ is the branching ratio of $l$ and $s_l(E)$ is the spectrum of branch $l$.

In a more realistic and complete reactor modelling exercise, contributions from non-fission-generated beta-unstable nuclides in the reactor fuel (\emph{i.e.} beta-decay of $^{239}$U $^{239}$Np bred into low-enriched fuel) or nearby materials (\emph{i.e.} neutron activation of structural aluminum in the core) must also be considered.
These contributions can be individually summed in a similar style as given in equation~\ref{eq:sum},
\begin{equation}
\label{eq:othersum}
    F_{other}(E, t) = \sum_k r_{k}(t)S_{k}(E),
\end{equation}
for nuclide $k$'s contribution to the total neutrino spectrum.  In this expression, $r(t)$ is the time-dependent beta-decay activity of nuclide $k$, while $S$ is its (potentially multi-branch) generated beta spectrum.  

As illustrated in figure~\ref{fig:sumblock}, the summation input can be user-supplied vectors of time dependent fission rates $\mathbf{R}(t)$ of all contributing isotopes, as 
\begin{equation}
    \mathbf{R}(t) = (R_{235}(t, E_i), R_{238}(t, E_i), ..., R_k(t,E_i)),
\end{equation}
where $E_i$ is the fission triggering incident neutron energy, and $R_k$ is the fission rate of fissile isotope $k$.
If non-fission (`nf') neutrino contributions are also being propagated in the calculation, the user also supplies a vector of decay rates of relevant nuclides $\mathbf{V}_{nf\beta}(t)$: 
\begin{equation}
    \mathbf{V}_{nf\beta}(t) = (r_1, r_2, ..., r_k).
\end{equation}
After identifying the relevant fission products from the nuclear database through the FPYEngine, CONFLUX generates from the user-specified $\mathbf{R}(t)$ the vectors $\mathbf{V}_{\beta}(t)$ and $\mathbf{V}_{ne\beta}(t)$, which are the time-dependent decay rates of fission produced beta-unstable isotopes and the non-equilibrium beta isotope contributions.  
The non-equilibrium isotope contribution to the reactor will rely on reactor information that contains the time dependent production of fissile isotopes and intermediate reactor products, which are reactor specific. 
Alternatively, CONFLUX has additional functionality to generate summation predictions using only user-specified $\mathbf{V}_{\beta}(t)$, $\mathbf{V}_{ne\beta}(t)$, and $\mathbf{V}_{nf\beta}(t)$.  
In this case, the user would require an upstream reactor simulation capable of generating predictions of individual fission product decay rates in the modeled core.  
Users can also perform summation calculations using only one of these components: in the case of using CONFLUX to study an individual beta-decaying nuclide's neutrino or beta emissions, the user would specify a one-component input $\mathbf{V}_{nf\beta}(t)$ vector.  

\begin{figure}
    \centering
    \includegraphics[width=0.7\linewidth]{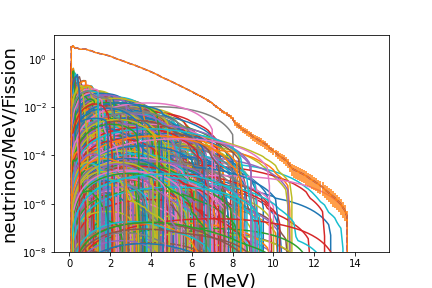}
    \caption{Equilibrium neutrino spectrum generated following thermal neutron fission of \ufive \, predicted using JEFF-3.3. cumulative FPY inputs.  Uncertainties include beta spectrum shape uncertainty of each contributing beta decay spectrum and averaged yield per fission.}
    \label{fig:sum_235}
\end{figure}

The beta spectrum engine within CONFLUX takes the list of beta isotopes specified within $\mathbf{V}_{\beta}(t)$, $\mathbf{V}_{ne\beta}(t)$, and $\mathbf{V}_{nf\beta}(t)$, searches CONFLUX's parsed ENSDF database for their beta decay information, including branching ratios, endpoint energies, atomic sizes, and etc. (detailed in section~\ref{sec:BSG}), to calculate the summed neutrino spectrum of each individual beta-unstable isotope, and to provide a vector of beta spectra $\mathbf{V}_{\beta}$, where
\begin{equation}
\label{eq:bigv}
    \mathbf{V_{\beta}} = (S_1(E), S_2(E), ..., S_k(E)).
\end{equation}
The $\mathbf{V_{\beta}}$ can either be a list of neutrino spectra beta spectra.
As last, the result of the summation calculation is shown as
\begin{equation}
    F(t, E) = (\mathbf{V}_{\beta}(t) + \mathbf{V}_{ne\beta}(t) + \mathbf{V}_{nf\beta}(t))\cdot \mathbf{V_{\beta}}.
\end{equation}
The summed equilibrium neutrino spectrum generated by thermal fissions of \ufive \, calculated using cumulative FPYs inputs from the JEFF-3.3 nuclear database, is shown in figure~\ref{fig:sum_235}.  
The versatile data-handling capability of CONFLUX is illustrated in figure~\ref{fig:sum_235} by the ability to easily visualize both the aggregate neutrino spectrum as well as the contribution of each individual fission product.  

\subsection{Use of Databases}
The CFY-DB, or the Cumulative Fission Yield Database, shown in figure~\ref{fig:sumblock}, are extracted from nuclear database, including ENDF/B-VIII~\cite{brown_endfb-viii0_2018} and JEFF-3.3~\cite{plompen_joint_2020}.
The nuclear databases used by CONFLUX are parsed into extensible markup language (xml) format for readability.
The default FPY databases are prepackaged and parsed as illusrated in in listing~\ref{lst:FPY}.
The root of databases consists of \texttt{HEAD}s that contain the atom mass factor \texttt{AWR}, atom number and mass combination \texttt{fissionZA} as isotope ID, number of fission modes in terms of incident neutron energy \texttt{LE}, and datatype \texttt{MT} being either independent fission products (\texttt{IFP}), or cumulative fission products (\texttt{CFP}).
Data of each fission mode is titled as \texttt{LIST}s, including incident neutron energy \texttt{Ei} that indicates spontaneous, thermal, fast neutron, and epithermal fission, as well as \texttt{NFPi}, the corresponding number of fission products.
The fission products are listed under each \texttt{LIST} as \texttt{CONT}s, storing information including isotope ID \texttt{ZA}, the atom number and mass combination, yield per fission \texttt{Y}, the absolute yield uncertainty \texttt{DY}, and isomeric state \texttt{FPS} (0 for ground states, 1 for isomeric states).
By default, the FPY Engine reads the cumulative fission yield, CFP, from the FPY databases.
In the cases of calculating neutrinos from reactors or fission from a non-equilibirium scenario, the independent FPYs can be read in the IFP database. 
When IFP is selected, users can provide a time window after each fission to calculate the evolution of neutrino flux in a non-equilibrium situation.
The non-equilibrium calculation cases will also adjust the rate of beta-decays in the model based on the half-lives of the beta-unstable isotopes.
The decay chains of the IFP beta decays are taken into consideration so that neutrinos from both the direct fission products and the beta-decays generations through the decay chain can also be calculated. 
The time dependent contribution fractions of the decay products in a decay chain can be expressed as:
\begin{equation}
    r_k(t) = \prod_i(1-e^{-\lambda_{i}t}),
\end{equation}
where $\lambda_i=\ln{(2)}/t_{\frac{1}{2}}$ is the decay constant of the $i$th generation of a decay chain began from the isotope $k$.
The fission yields from each isotope are multiplied by its total fuel contribution, most commonly in the format of total number of fissions, then summed to build $\mathbf{V_{\beta}}(t)$.
\begin{lstlisting}[caption={Example of the parsed FPY database in xml format.},captionpos=t, label=lst:FPY]
<?xml version="1.0" ?>
<nfy-092_U_235>
	<HEAD AWR="233.025" FissionZA="92235" LE="3" MT="IFP">
		<LIST Ei="0.0253" Ii="2" NFPi="1247">
			<CONT DY="1.3122e-19" FPS="0.0" Y="2.05032e-19" ZA="23066"/>
			<CONT DY="1.54228e-14" FPS="0.0" Y="2.40981e-14" ZA="24066"/>
			...
	<HEAD AWR="233.025" FissionZA="92235" LE="3" MT="CFP">
		<LIST Ei="0.0253" Ii="2" NFPi="1247">
			<CONT DY="1.3122e-19" FPS="0.0" Y="2.05032e-19" ZA="23066"/>
			<CONT DY="1.54229e-14" FPS="0.0" Y="2.40983e-14" ZA="24066"/>
			...
\end{lstlisting}

The beta-DB that provides the beta decay information of each isotope is parsed from ENSDF~\cite{TULI1996506}.
As shown in listing~\ref{lst:betaDB}, the beta database consists of lists titled with \texttt{isotope} that contains the isotope name, the identity "ZAI" \texttt{isotope= Z*1000 + A*10 + I}, where I is the isomeric state (0 or 1), the Q value, and the half life \texttt{HL}.
Each listed beta branch contains the branch fraction and uncertainty, the end point energy $E0$ and uncertainty, as well as the spin and parity difference $\Delta J\pi$ between the decay parent and daughter, where the spin difference is noted as an integer and parity difference is noted as ``-'' if there is parity change.
Isotopes with missing branch information are left as a blank list. 
The beta spectrum engine reads a list of isotopes and with the decay information to calculate the list of beta/neutrino spectra $\mathbf{V_{\beta}}$.
\begin{lstlisting}[caption={Example of the parsed beta database in xml format.},captionpos=t, label=lst:betaDB]
<?xml version="1.0" ?>
<betaDB>
    ...
	<isotope name="Ar41" isotope="180410" Q="2.492" HL=" 109.61 M ">
		<branch fraction="0.008" sigma_frac="0.000" end_point_E="2.492" sigma_E0="0.000" dJpi="2-"/>
		<branch fraction="0.992" sigma_frac="0.000" end_point_E="1.198" sigma_E0="0.000" dJpi="0"/>
		<branch fraction="0.001" sigma_frac="0.000" end_point_E="0.815" sigma_E0="0.001" dJpi="0-"/>
	</isotope>
	<isotope name="P42" isotope="150420" Q="18.65" HL="48.5 MS   "/>
	<isotope name="S42" isotope="160420" Q="7.28" HL="1.016 S   ">
		<branch fraction="0.176" sigma_frac="0.011" end_point_E="6.013" sigma_E0="0.140" dJpi="1"/>
		<branch fraction="0.007" sigma_frac="0.002" end_point_E="5.894" sigma_E0="0.140" dJpi="0"/>
		<branch fraction="0.016" sigma_frac="0.008" end_point_E="5.569" sigma_E0="0.140" dJpi="0"/>
		...
	</isotope>
	...
\end{lstlisting}

Default fission databases are prepackaged in the \texttt{CONFLUX/data} folder along with the Python parser programs that convert the nuclear data from the original formats to the CONFLUX xml formats.  
The latter allows for generation of updated CONFLUX databases in the event of new nuclear data tabulations, and for generation of user-customized databases.

\subsection{Uncertainty calculations}
\begin{figure}
    \centering
    \includegraphics[width=0.7\linewidth]{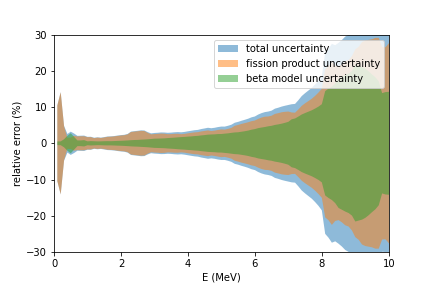}
    \caption{Contributions of different types of uncertainties relative to the \ufive \, neutrino spectrum calculated based on the JEFF-3.3 fission product database and ENSDF (2024.07.10 ver.) beta decay data, as an example of the CONFLUX calculated reactor neutrino uncertainty.}
    \label{fig:sum_uncertainties}
\end{figure}

The uncertainties of the summation mode calculation include the fissile isotope fraction uncertainty from the user-inputted reactor model, the uncertainty of individual beta spectrum, the beta branch fraction uncertainty, and the uncertainty of FPY.
The fractional neutrino spectrum uncertainties the previously-shown example of equilibrium thermal-neutron-induced \ufive~fission are shown in figure~\ref{fig:sum_uncertainties}.  
Uncertainties from the input reactor model are expected from the users, which can be provided either as independent uncertainties of fissile isotope rates, or as an uncertainty covariance matrix among the provided fissile isotopes.  
When building a reactor model, each contributing fissile isotope can be added with an uncertainty of isotope fraction in the function \texttt{summation\_model.AddFissionIstp(U235, count = 100, d\_count = 5)}.
The contribution from fissile and non-fissile isotopes in a reactor can be provided in terms of either fractional uncertainties or uncertainties in the absolute amount of decays.  
If not provided by the user, the uncertainty in the fissile isotope contribution is set to zero by default.

The beta spectrum uncertainty can result from a series of uncertain factors, e.g., error in the end-point energy from nuclear database, theory uncertainties in the calculation of forbidden transition corrections, and uncertainties in higher order theoretical corrections.
Full assessment of all factors of beta-decay-related spectrum uncertainties are difficult to provide, given the abundance of missing information in nuclear structure and fission product databases.
However, specific error sources, such as end-point energy uncertainties, can be simply propagated in the spectrum calculation.  
For this example uncertainty source, errors are determined in CONFLUX as each beta or neutrino spectrum is calculated repeatedly through MC sampling with varying endpoint energies within the uncertainty interval. 
By assuming that end-point energy uncertainties of beta branches recorded in ENSDF are correct 1$\sigma$ interval, the MC spectrum samples are generated with randomized end-point energy from a Gaussian distribution.
The standard deviation of all MC samples in each energy bin is characterized as the default beta spectrum uncertainty $\delta s(E)$.
Other uncertain factors will rely on user input.  
 
The uncertainty of beta feedings are stored in the beta-DB. 
If one assumes uncorrelated beta branching uncertainties, the summed neutrino spectrum error is simply the summed single branch spectrum uncertainty weighed by the uncertainty of branch fractions, as
\begin{equation}
\label{eq:nocorr}
    \sigma_S(E) = \sum_{i}b_i s_i(E)\sqrt{\left(\frac{\sigma b_i}{b_i}\right)^2+\left(\frac{\sigma s_i(E)}{s_i(E)}\right)^2}.
\end{equation}
However, the branching fraction can be highly correlated among branches of single isotopes. 
While the calculation with complete covariance needs explicit studies about the experimental measurements behind each isotope, it is easy to implement the anti-correlation between the ground state decay and other branches (commonly referred as Ground State Feeding~\cite{PhysRevC.102.064304}), as
\begin{equation}
\mathbf{R}_{branch} = 
\begin{pmatrix}
1 & -1/b_1 & -1/b_2 &\hdots & -1/b_i\\ 
-1/b_1 & 1 & 0 & \hdots & 0\\
-1/b_2 & 0 & 1 & \hdots & 0\\
\vdots & \vdots & \vdots & \ddots & \vdots \\
-1/b_i & 0 & 0 & \hdots & 1
\end{pmatrix},
\end{equation}
where the first row represents the ground state decay branch, and $b_i$ is the fraction of branch $i$.
Thus, the uncertainty of multi-branch beta decay with covariance among branches can be calculated through
\begin{equation}
    \mathbf{\Sigma}_{branch} = \mathbf{\Delta}\mathbf{R}_{branch}\mathbf{\Delta^T}.
\end{equation}
where $\mathbf{\Delta}$ is the vector of branch uncertainties.
Users can also provide customized covariance matrix through the method~\texttt{BetaEngine.EditIsotope(ZAI, betabranchlist, cov)}.
The covariance matrix is then propagated to the summed uncertainty of an isotope, as
\begin{equation}
    \sigma^2_S(E) = \sum_{i}\sigma^2(b_i s_i(E)). 
\end{equation}

The uncertainty of FPY, extracted from the ENDF or JEFF databases, are described in terms of an uncertainty for each fission product.  
In other words, uncertainty information stored in the databases are individually recorded, despite the presence of potential experimental measurement correlations between nuclides.  
FPY uncertainty propagation can be processed without covariance similarly as equation~\ref{eq:nocorr}, as
\begin{equation}
    \sigma_S(E) = \sum_{i}f_i s_i(E)\sqrt{\left(\frac{\sigma f_i}{f_i}\right)^2+\left(\frac{\sigma S_i(E)}{S_i(E)}\right)^2}.
\end{equation}
When correlation or covariance matrices of fission products are provided in the calculation, the total uncertainty of a fission isotope neutrino spectrum is calculated with 
\begin{equation}
    \mathbf{\Sigma}_{FPY} = \mathbf{\Delta}_{FPY}\mathbf{R}_{FPY}\mathbf{\Delta^T}_{FPY}.
\end{equation}
Prepackaged example covariance and correlation matrices are included for the fission products from  \ufive, \ueight, \punine, and \puone, calculated by the work referred in~\cite{matthews_stochastically_2021}, a MC calculation based on ENDF.B/VIII and JEFF-3.3.
CONFLUX prepackages a python script that downloads and parses these covariance and correlation matrix data.  
The matrices are 2D array formatted in csv, with indices being the fission products ZAI.
The users need only to provide the customized covariance matrices among selected isotopes and fission products in the same format for more customized uncertainty calculations.  

\subsection{Assumptions in the default summation}
\begin{figure}
    \centering
    \includegraphics[width=0.5\linewidth]{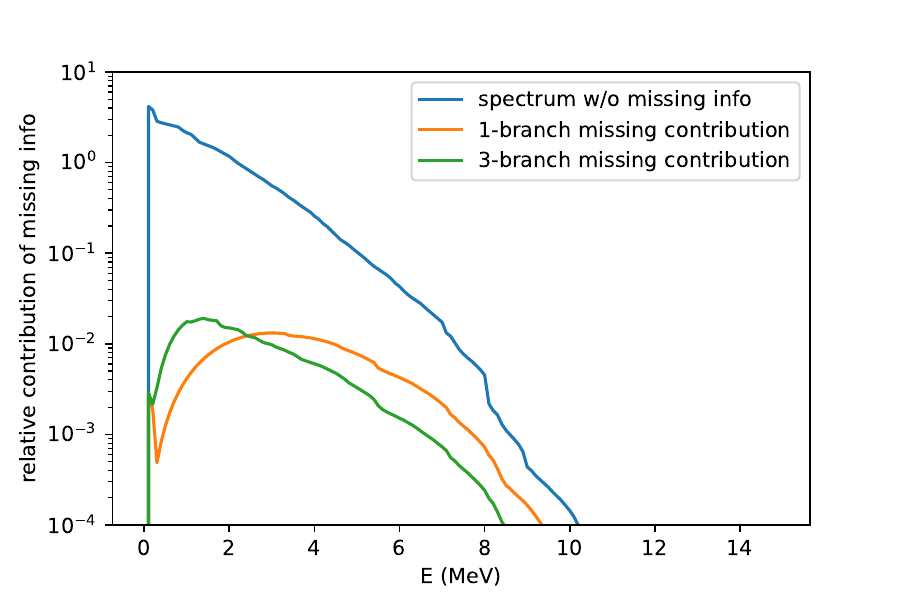}
    \caption{Illustration of assumptions applied to info-missing isotopes. Cases included are no missing isotopes added versus the missing isotope summed neutrino spectrum with the default three-branch decay mode, and a simplified single ground state decay mode.}
    \label{fig:missing}
\end{figure}

There are unknown information about fission products and beta branches in the databases. 
In ENSDF, there exist beta-unstable isotopes with no decay information.
To allow user to add contributions from the missing decays, an optional argument in the BetaEngine contain a beta spectrum is calculated by assuming the isotopes with missing info decay all through three branches with same fractions, where the end-point energies equal 1/3, 2/3 and total Q value, i.e. $E_0=Q/3$, or $2Q/3$, or $Q$. 
The number of branches and the end-point energies of info-missing isotopes can be individually customized to let users input the details of their individual isotopes studies (see figure~\ref{fig:missing}).
There are also beta-branches with multiple spin-parity changes or missing information about spin-parity change.
The former are assumed to decay through the most allowed transition, and the later is assumed to be allowed decay.
One can modify these assumption by choosing whether to calculate beta spectra with missing information or editing specific isotope with their own information.

\subsection{Practice example}
\begin{lstlisting}[caption={Example of a full summation process.},captionpos=t, label=lst:sumexample]
# conflux modules
from conflux.BetaEngine import BetaEngine
from conflux.FPYEngine import FissionIstp
from conflux.SumEngine import SumEngine

# generate a spectrum dictionary by through theoretical beta calculation
betaSpectraDB = BetaEngine(xbins=np.arange(0, 15, 0.1))
betaSpectraDB.CalcBetaSpectra(nu_spectrum=True, branchErange=[-1.0, 20.0])

# initialize fission isotopes
U235 = FissionIstp(92, 235, Ei=0)
U235.LoadFissionDB(defaultDB='ENDF')
U235.LoadCorrelation(defaultDB='ENDF')
U235.CalcBetasSpectra(betaSpectraDB)

U238 = FissionIstp(92, 235, Ei=0.5)
U238.LoadFissionDB(defaultDB='JEFF')
U238.LoadCorrelation(defaultDB='JEFF')
U238.CalcBetasSpectra(betaSpectraDB)

# summing beta/neutrino spectra with respect to the counting of fissions
summation_model = SumEngine(betaSpectraDB)
summation_model.AddFissionIstp(U235, "U235_thermal", count = 100, d_count = 5)
summation_model.AddFissionIstp(U238, "U238_fast", count = 100, d_count = 5)
summation_model.CalcReactorSpectrum()

# result spectrum and uncertainty
summed_spect = summation_model.spectrum
summed_err = summation_model.uncertainty
\end{lstlisting}

\subsection{Verification of calculation results}
The beta decay spectrum calculated by BSG~\cite{hayen_beta_2019} has already been tested and recognized by the field.
Thus, the verification of calculated results from CONFLUX has focused on proving that the calculations involving covariance matrices are mathematically correct. 
An artificial covariance matrix among three fission products were defined and used to calculate a three-product fission yield model.  
The total spectrum uncertainty of such an artificial fission model is calculated separately from CONFLUX and was then compared to the uncertainty calculated by CONFLUX using the same customized FPY and covariance, and shown the be mathematically identical.  


\section{Beta Conversion Calculation}
\label{sec:b-conversion}
\begin{figure}[h]
    \centering
    \includegraphics[width=0.9\textwidth]{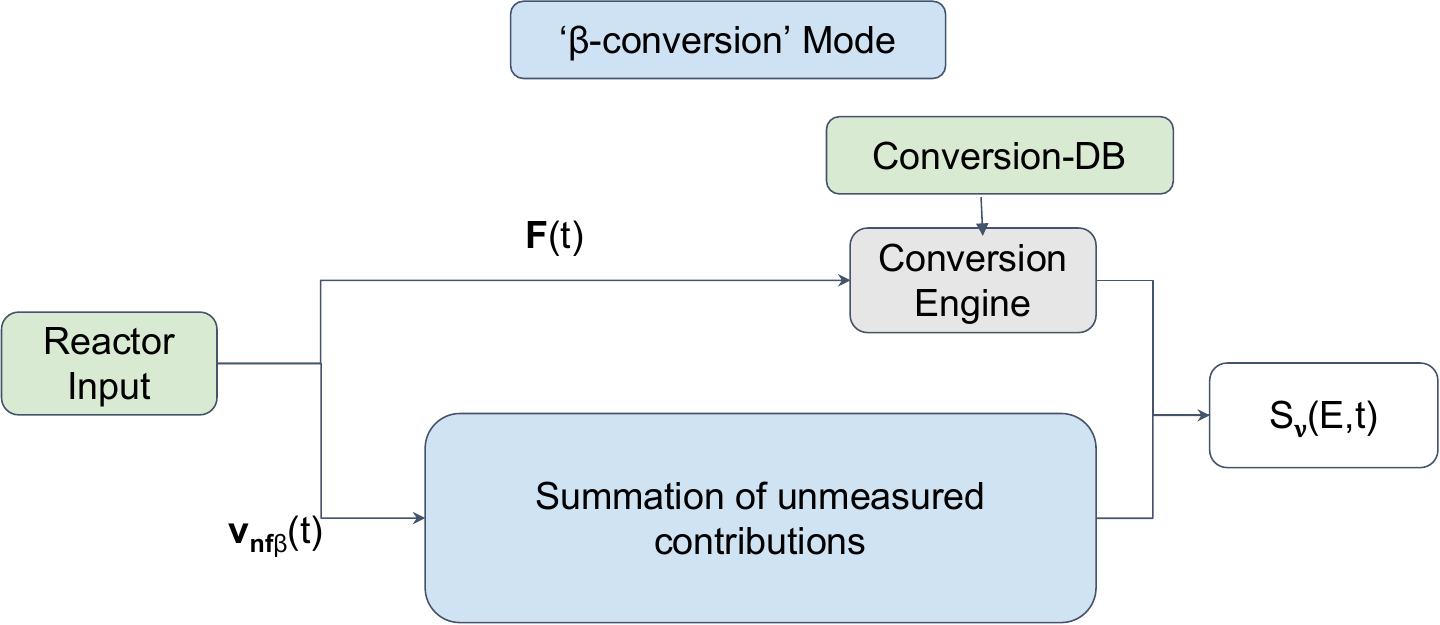}
    \caption{Block diagrams of the $\beta$-conversion mode.}
    \label{fig:convblock}
\end{figure}

As introduced in section~\ref{sec:general}, the neutrino spectrum generated in a reactor by beta-unstable fission products can also be calculated by applying the principle of energy conservation to direct measurements of aggregate fission beta spectra~\cite{huber_determination_2011, mueller_improved_2011}.  
Conservation of energy and momentum enables direct conversion between the fission-generated neutrino spectrum and beta spectrum, providing a more direct, data-driven approach for neutrino flux prediction compared to the summation calculation.  
In CONFLUX, the required factors for a beta conversion calculation include the beta spectrum data collected in a series of ILL-based measurements~\cite{buhring_screening_1984, hahn_antineutrino_1989, von_feilitzsch_experimental_1982} and a conversion calculation engine that determines  a best fit beta spectra to be converted to neutrino spectra, as shown in figure~\ref{fig:convblock}.  
Non-fissile contributions and non-equilibrium effects were not considered in the collection of the reference aggregate fission beta spectra used for conversion calculations.
Thus, calculations of neutrino contributions for these categories rely on the summation-mode calculation described in section~\ref{sec:summation}.  

\subsection{Beta Spectra of Fissile Isotopes}
The most commonly used aggregate beta spectra used in conversion calculations, described in~\cite{von_feilitzsch_experimental_1982, hahn_antineutrino_1989}, are henceforth referred to as the ILL measurements.
Beta spectra from fission of \ufive, \punine, and \puone~are measured, while the spectrum of \ueight~was not measured, meaning that neutrino contributions for the latter isotope also must rely on the summation engine.  
Recently, \ufive \, and \punine \, aggregate beta spectra were remeasured as described in~\cite{PhysRevD.104.L071301}; given the differential nature of these measurements, these data are currently not included as default reference data for CONFLUX.  
The finest binning for the energy spectra is 50 keV, referred from the re-published data~\cite{haag2014republicationdatamagneticspectrometer}, and then prepackaged in CONFLUX in .csv format.

\subsection{Virtual Branches}
Aggregate fission beta spectra are the result of the summed contributions from spectra of many individual fission products.  
It is unrealistic to fit the contributions from the thousands of existing beta branches to the summed beta spectra measured from fissile isotopes.  
Therefore, the reference fission beta spectra were fitted with a limited number of virtual branches whose end-point energies are arithmetical in the fitted energy range.
Contributions and end-point energies are randomized to search for a best fit summed beta spectrum compared to the reference spectrum.  

First, the virtual beta spectra are generated through the CONFLUX theoretical beta spectrum generator function, e.g. the beta engine used to generate $\mathbf{V_{\beta}}$ described in equation~\ref{eq:bigv}.  
From the highest energy to the lowest, slices of the reference beta spectrum are fitted with the virtual spectra.
Isotopes whose end-point energies are within the energy range of each spectrum slice are searched for in the FPY and beta database.  
These isotopes are grouped together if they contain beta-decay branches with end-point energies in the corresponding energy range.  
The average atom number $Z$ and atom mass $A$ of all beta-decay branches in the target energy range is used as the $Z$ and $A$ number for the virtual branch for the fitted spectrum slice.  
All virtual branches are set to used allowed transition spectrum shapes by default to simplify the calculation procedure, while an optional calculation method exists to tally the percentage of forbidden transition contributions among all beta branches in the corresponding energy range.  
In the latter case, the virtual spectrum shape would be composed an allowed spectrum and forbidden spectrum based on the tallied result.  
For each virtual branch, the end point energy $E_0$ and branch contribution to the total spectrum $f_i$ are randomized within a given range to search for the best-fit virtual branch compared to the reference beta spectrum data. 
The fitting procedure follows these steps: (1) the virtual branch spectrum of the highest end-point energy is fitted to the spectrum slice in the highest energy range; (2) once the best-fit beta spectrum of the target energy range is found, the virtual spectrum is subtracted from the reference spectrum; (3) the virtual branch of the subsequent spectrum slice is fitted to the remainder of the subtracted reference spectrum in the corresponding energy range; (4) the steps continue until the reference spectrum is completely subtracted by the best-fit virtual spectra. 
The contribution (or normalization) of each virtual spectrum and the best-fit end-point energy are recorded to build the beta spectrum model for conversion.
The best fit beta spectrum is the sum of best fit beta spectra, written as
\begin{equation}
    S_{\beta}(E) = \sum_i f_iS^{E_0}_{\beta i}(E).
\end{equation}
Then, the list of best fit end-point energy $E_0$ and contribution $f_i$ are used to generate a list neutrino spectra through the direct conversion from virtual beta spectra to a summed neutrino spectrum.  

\subsection{Uncertainty Calculation}
The uncertainty of the conversion calculation is driven by the uncertainties in the reference beta data, i.e. the uncertainty of the original fission beta spectrum measurements.
If the user provides new beta spectrum data, uncertainties of the measurement will be needed in the same .csv format.
To calculate the uncertainty of the converted reactor neutrino spectrum, CONFLUX generates Monte Carlo toy reference spectra with randomized bin contents within the uncertainty interval of the referenced data.  
The randomization of the spectrum ignores correlations between energy bins.
The summed best-fit virtual spectra of all toys are compared to the best-fit beta spectrum of the actual data to generate a covariance matrix as the uncertainty fitted beta spectra.  
The Monte Carlo best fit beta spectra are converted to the neutrino spectrum to calculate the covariance matrix and uncertainty of the converted neutrino spectrum result.  

\subsection{Practice example}
An example of the conversion calculation is shown listing~\ref{lst:convexample}.
\begin{lstlisting}[caption={Example of a full conversion process.},captionpos=t, label=lst:convexample]
# conflux modules
from conflux.BetaEngine import BetaEngine, BetaBranch
from conflux.FPYEngine import FissionModel, FissionIstp
from conflux.ConversionEngine import ConversionEngine, BetaData

# initialize the beta spectrum data to be fitted against
beta235 = BetaData("./data/conversionDB/U_235_e_2014.csv")
beta239 = BetaData("./data/conversionDB/Pu_239_e_2014.csv")

# initialize fission istoopes.
U235 = FissionIstp(92, 235, Ei=0)
Pu239 = FissionIstp(94, 239, Ei=0)
# load fission product yields to calculate average atom number
U235.LoadFissionDB()
Pu239.LoadFissionDB()

# fit the beta spectrum, each fission isotope is provded with a fission fraction
convertmodel = ConversionEngine()
convertmodel.AddBetaData(beta235, U235, "U235", 0.7, 0.01)
convertmodel.AddBetaData(beta239, Pu239, "Pu239", 0.3, 0.01)
convertmodel.VBfit(slicesize=0.25)

# calcualte the covariance matrix of the virtual branch fitting
xval = np.linspace(0., 10., 200)
covmat235=convertmodel.vblist["U235"].Covariance(beta235, xval, nu_spectrum = False, samples=50)
covmat239=convertmodel.vblist["Pu239"].Covariance(beta239, xval, nu_spectrum = False, samples=50)

# summing two spectra with respect to the fission fraction
spectrum239 = convertmodel.vblist["Pu239"].SumBranches(xval, nu_spectrum = False)
spectrum239Nu = convertmodel.vblist["Pu239"].SumBranches(xval, nu_spectrum = True)

\end{lstlisting}

\subsection{Verifying the calculation result}
To verify the accuracy of the conversion prediction, a pair of synthetic beta and neutrino spectra were generated using the CONFLUX summation mode. 
The synthetic data were generated with the same binning of the ILL legacy reference beta spectrum data.
In the range from 2 to 8 MeV, the best fit beta spectrum differs less than 0.1\% compared to the synthetic beta spectrum. 
After converted to neutrino spectrum, the difference between the converted spectrum and the synthetic neutrino spectrum contains 2.5\% variations on average, which covers the absolute difference in the 2 to 8 MeV range. 
The spectral difference is smaller than the corresponding fitting uncertainty.
The total neutrino flux difference between synthetic and converted cases is approximately 0.4\% in the same range.  
\begin{figure}
\centering
    \includegraphics[width=.48\linewidth]{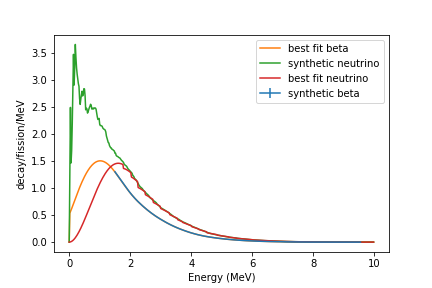}
    \includegraphics[width=.48\linewidth]{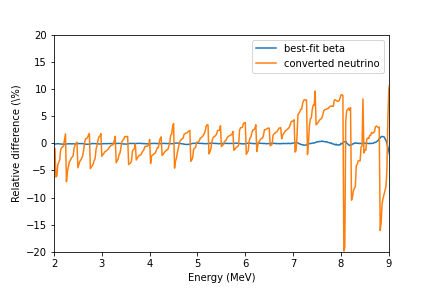}
\caption{Verification of the results of the conversion mode calculation. Left: the synthetic spectra and best fit spectra. Right: relative differences between the synthetic beta and best fit beta, as well as the relative difference between the synthetic neutrino spectrum to the converted neutrino spectrum. The saw teeth feature is the result of discontinuity of the virtual neutrino spectrum end-points. }
\label{fig:convert_verify}
\end{figure}

\section{Research Examples}
\label{sec:example}
CONFLUX is developed to be flexible and enable a variety of use cases across a range of research topics.
Potential applications of CONFLUX include time-dependent reactor neutrino modeling for fundamental and applied physics purposes, antineutrino source term modeling for other radiations sources such as fission pulses or spent nuclear fuel, and theoretical beta spectrum calculations with modified or customized data and beta decay theory inputs for the purposes of improving understanding of reactor flux modeling uncertainties or generating predictions for sterile neutrino mass measurement experiments.
This section presents use case concepts in the form of a series of highlighted examples.  
We encourage readers and users to be creative and communicative in their consideration of the potential applications of  CONFLUX.

\subsection{Summation calculation with reactor fissile composition}
To calculate the neutrino flux of a specific reactor, one can provide to CONFLUX time-dependent fission rates for the primary fission fission isotopes as described in section~\ref{sec:general}.  
We envision this being one of the most common inputs provided by a user for summation calculations using CONFLUX.  
In this example, we provided the time-dependent reactor model shown in table~\ref{tab:timedepend} as an input for such a calculation.  
The model shown, taken from~\cite{giunti_reactor_2017}, corresponds to the case of a 3~GW$_{\textrm{th}}$ commercial low-enriched uranium (LEU) reactor's fission fragment evolution in the days following reactor start-up, ignoring non-equilibrium and non-fissile isotopes.  
The code for running this example can be found in \texttt{examples/TimeEvolvingReactor.py}.  

\begin{table}[!htp]\centering
\caption{Fission rate of four major fissile isotopes in a LEU reactor with respect of time.}\label{tab:timedepend}
\scriptsize
\begin{tabular}{lrrrrr}\toprule
&Fission rate & & & \\
Time (days) &U235 &U238 &Pu239 &Pu241 \\\midrule
0 & $1.31\times10^{19}$ & $7.79\times10^{17}$ &$3.69\times10^{-15}$ &$3.64\times10^{-15}$ \\
0.1 &$1.31\times10^{19}$ &$7.80\times10^{17}$ &$1.75\times10^{13}$ &0.00 \\
2 &$1.30\times10^{19}$ &$8.02\times10^{17}$ &$8.94\times10^{15}$ &$0.00$ \\
20 &$1.27\times10^{19}$ &$8.05\times10^{17}$ &$3.04\times10^{17}$ &$1.75\times10^{14}$ \\
100 &$1.16\times10^{19}$ &$7.96\times10^{17}$ &$1.41\times10^{18}$ &$1.82\times10^{16}$ \\
200 &$1.04\times10^{19}$ &$8.11\times10^{17}$ &$2.38\times10^{18}$ &$9.83\times10^{16}$ \\
300 &$9.50\times10^{18}$ &$8.29\times10^{17}$ &$3.11\times10^{18}$ &$2.32\times10^{17}$ \\
400 &$8.68\times10^{18}$ &$8.56\times10^{17}$ &$3.68\times10^{18}$ &$4.03\times10^{17}$ \\
600 &$7.20\times10^{18}$ &$9.11\times10^{17}$ &$4.63\times10^{18}$ &$8.21\times10^{17}$ \\
800 &$5.84\times10^{18}$ &$9.72\times10^{17}$ &$5.39\times10^{18}$ &$1.28\times10^{18}$ \\
1,000.00 &$4.58\times10^{18}$ &$1.04\times10^{18}$ &$6.06\times10^{18}$ &$1.75\times10^{18}$ \\
1,200.00 &$3.41\times10^{18}$ &$1.10\times10^{18}$ &$6.66\times10^{18}$ &$2.19\times10^{18}$ \\
1,400.00 &$2.38\times10^{18}$ &$1.16\times10^{18}$ &$7.20\times10^{18}$ &$2.57\times10^{18}$ \\
1,600.00 &$1.54\times10^{18}$ &$1.21\times10^{18}$ &$7.62\times10^{18}$ &$2.88\times10^{18}$ \\
\bottomrule
\end{tabular}
\end{table}

\begin{figure}[H]
    \centering
    \includegraphics[width=0.7\textwidth]{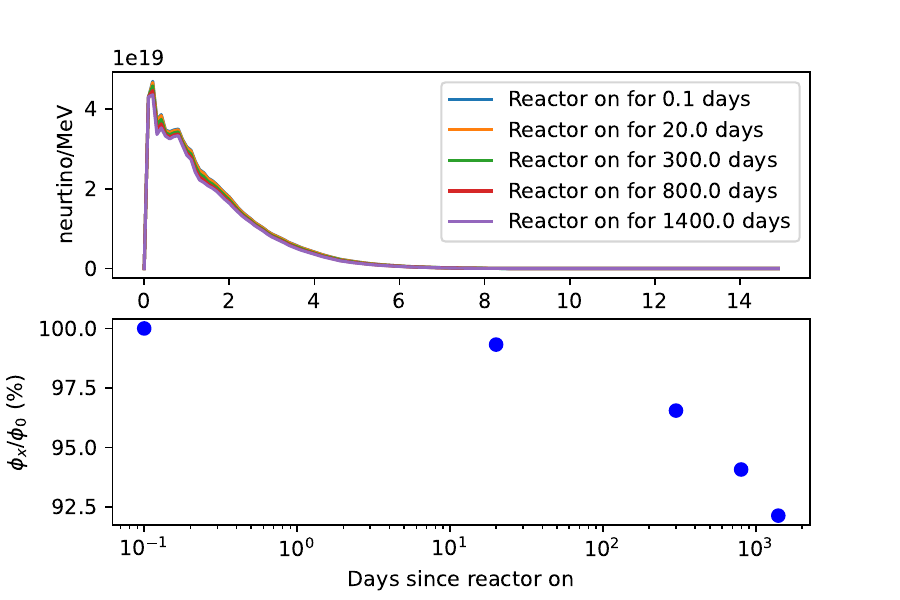}
    \caption{Top: The neutrino spectrum produced from fission by a 3~GW$_{\textrm{th}}$ LEU reactor at specified time periods after reactor start-up. Bottom: The energy-integrated neutrino flux generated by this reactor relative to the production rate at reactor start-up.  Cumulative FPYs are used for this summation example, meaning that all fission product beta decays are assumed to be in equilibrium with the rates of fission in the core at all times following reactor start-up.}
    \label{fig:sum_time}
\end{figure}

As described in the previous sections, at the beginning of the calculation, the user defines fissile isotopes  contributing fission, and CONFLUX then calculates the corresponding beta or antineutrino spectrum using BSG and the FPY from a specified or from the default nuclear databases. 
As seen in table~\ref{tab:timedepend}, for each time period, the user is providing the amount of fissions each isotope contributed in the reactor per second in the marked time period.  
The evolution of the fission-derived neutrino flux over time is made accordingly by CONFLUX as shown in figure~\ref{fig:sum_time}.  
The predicted decrease in emitted flux versus time after start-up is an outcome of the comparatively low neutrino yield of \punine, which is bred into the fuel as the cycle progresses.  

Cumulative FPYs are used for this summation example, meaning that all fission product beta decays are assumed to be in equilibrium with the rates of fission in the core at all times following reactor start-up.  
A more detailed reactor simulation would also incorporate contributions from beta-unstable isotopes not generated via fission, as well as non-equilibrium effects related to long-lived fission products that primarily manifest early in the reactor cycle; as described in the previous sections, CONFLUX is also capable of modeling both of these second-order effects.  

\begin{figure}[H]
    \centering
    \includegraphics[width=0.7\textwidth]{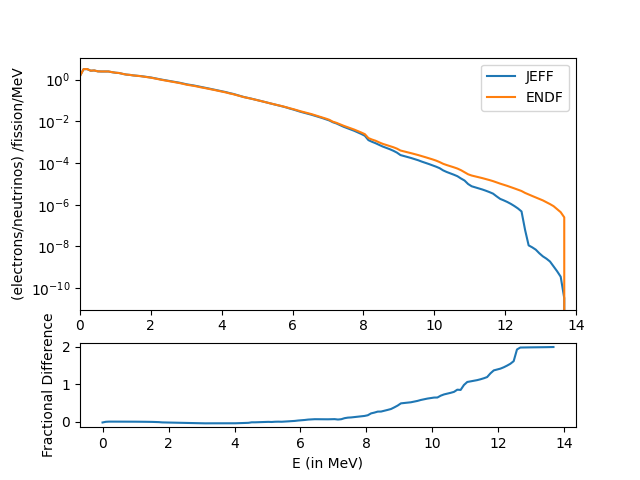}
    \caption{Comparison between neutrino spectra predicted with FPY of 235U from JEFF 3.3 and ENDF/b-VIII.0.}
    \label{fig:sum_crossDB}
\end{figure}

In CONFLUX, the JEFF-3.3~\cite{plompen_joint_2020} and ENDF/B-VIII~\cite{brown_endfb-viii0_2018} FPY databases are stored by default, allowing the user to switch the input FPY database used in a customized calculation.
Cross-database comparisons of neutrino flux can be made by changing the loaded database after the corresponding fissile isotope is defined. 
An example of a \ce{^235U}~neutrino spectrum cross-database comparison is stored in CONFLUX at {\texttt{examples/SummationDBchanging.py}}.  
the JEFF vs ENDF output are shown in figure~\ref{fig:sum_crossDB}.  
User-specified FPY databases for other fission isotopes or database versions can also be generated with the prepackaged parser for databases in ENDF-6 format.   

\subsection{Summation calculation with a list of beta-unstable isotopes}
In addition to summing the neutrino flux contributions of fission products, CONFLUX can take a list individual beta-unstable nuclides for a summation calculation, as described in section~\ref{sec:summation}.  
A list of beta-unstable fission products with corresponding contributions and uncertainties one potential output of some reactor simulation toolkits. 
On the other hand, individual beta decay nuclides in the provided list can also be completely unrelated to the fission process.  

\begin{figure}[H]
    \centering
    \includegraphics[width=0.6\textwidth]{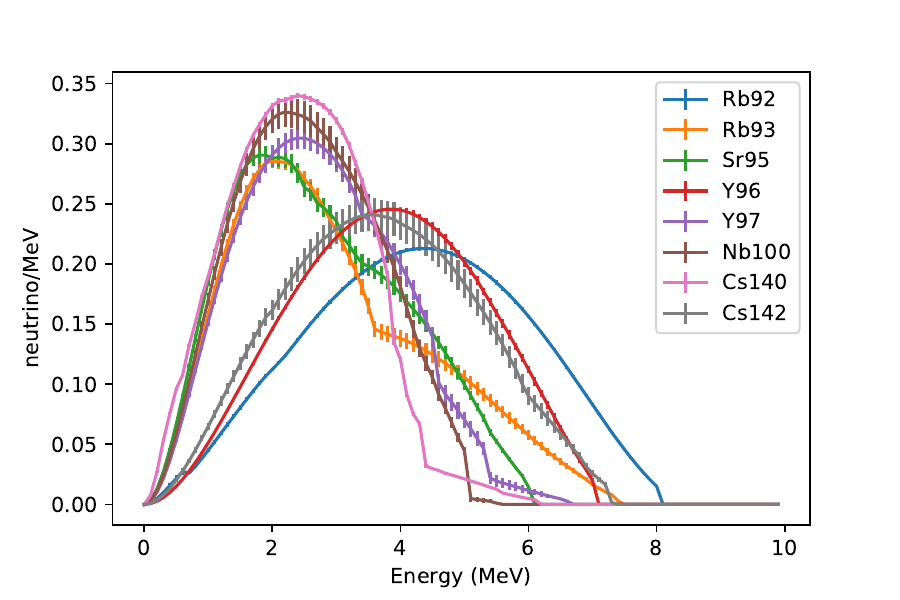}
    \caption{Neutrino spectrum per decay for selected fission products with both high \ce{^235U}~CFY and high beta-decay Q-values.  Displayed nuclides were chosen using a separate CONFLUX fission-produced reactor neutrino summation calculation.  Uncertainties associated with each nuclide's spectrum are also pictured.}
    \label{fig:sum_indie}
\end{figure}

The example at {\texttt{examples/selectedneutrinoSpec.py}} in CONFLUX is provided to guide the user in using a specified list of beta-unstable isotopes to calculate neutrino flux contributions and uncertainties via the summation method.  
The example performs a few iterations of the summation calculation, with, in each case, a single input beta-decay nuclide specified to have a decay rate of 1.  
Chosen nuclides are those possessing both high CFYs and high Q-values, which dominate the neutrino emissions of a reactor in the 5-7~MeV neutrino energy regime where existing neutrino measurements and summation/conversion predictions diverge from one another.   
We note that these nuclides were identified with a relatively minor extension of the previous provided example.  
Resulting neutrino spectra for each of the example nuclides are shown in figure~\ref{fig:sum_indie}.  
Plotted spectra also include propagated uncertainty bands.  

This process and example may be useful to users interested in exploring nuclides with interesting beta-decay end-point energies or transition types, high FPY or isomeric ratio uncertainties, or missing nuclear data, for the purposes of testing the uncertainty contribution or theoretical treatment of a group of isotopes involved in generating neutrino emissions.  
For precise reactor models that contain both fission fragments and a list of individual beta-unstable isotopes, CONFLUX allows the summation of two type of neutrino flux in this way.  

\subsection{Beta-decay customization}
\begin{figure}[H]
    \centering
    \includegraphics[width=0.49\textwidth]{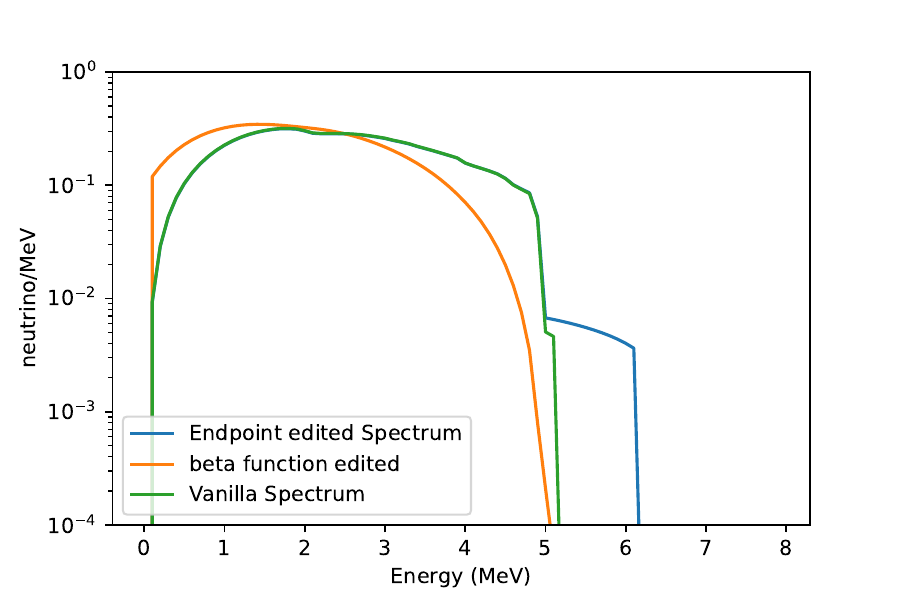}
    \includegraphics[width=0.49\textwidth]{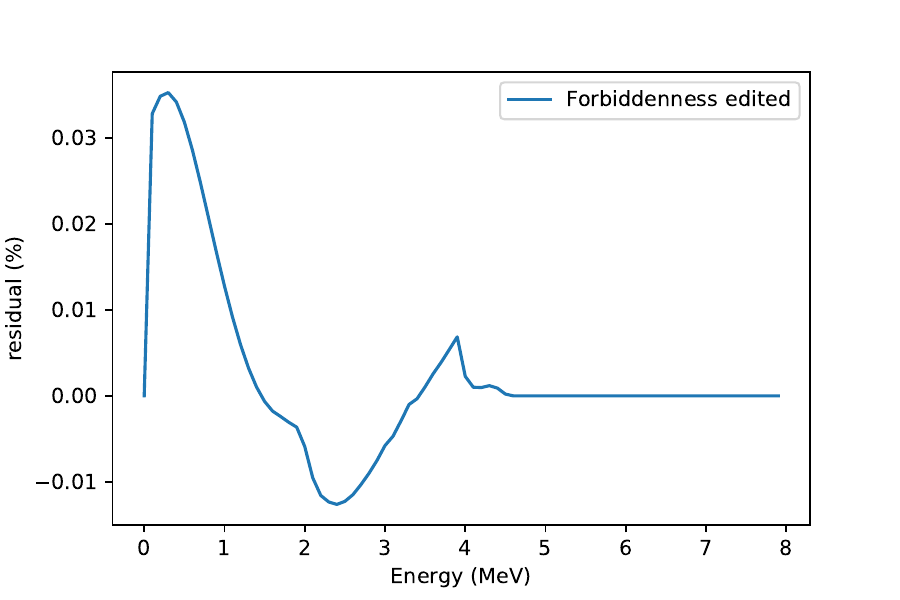}
    \caption{Neutrino spectrum of \ce{^241Pu} with edited beta decay properties. Left: Neutrino spectra with changed end-point energy and customized beta-decay function (pure fermi function in this case). Right: Residual between the default neutrino spectrum to a spectrum where allowed transitions are changed to the first unique forbidden transitions.}
    \label{fig:sum_change_input}
\end{figure}

A key feature of CONFLUX is that it allows customization of the calculation of emitted beta and neutrino spectra. 
The beta-unstable isotopes can be modified by editing the branching ratio, end-point energy, and type of allowed or forbidden transition. 
For more theoretical corrections outside of the default BSG beta-decay theory, CONFLUX allows the user to provide a custom function to the beta/neutrino spectrum generator while keeping other summation factors, such as FPYs, unchanged.  
The example at {\texttt{examples/BetaChangingParams.py}} indicates how the aggregate \puone~neutrino spectra is impacted by editing the properties of decay of its beta isotopes.  
Neutrino spectrum calculation results, shown in figure~\ref{fig:sum_change_input}, are given for the default case, a case where the end-point energy of a single fission product, \ce{^141Cs}, was altered, and a case where the forbiddenness of all fission product beta decays was altered from allowed to the first order unique forbidden transition.  
This demonstrated feature of CONFLUX can be vital to studies looking to easily improve modeling of beta-decays for nuclear physics experiments, as well as to studies seeking to understand the impacts of new or updated inputs to reactor neutrino predictions.

\begin{figure}[H]
    \centering
    \includegraphics[width=0.6\textwidth]{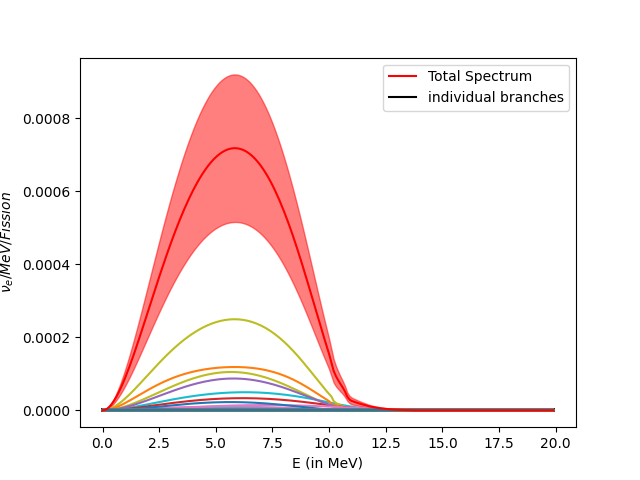}
    \caption{Thermal \ce{^235U}~fission neutrino spectra from individual fission products beta decay branches whose Q value is greater 10~MeV. The red line is the total spectrum for of all branches whose Q value $>$ 10~MeV. CONFLUX-propagated summation calculation uncertainties in the aggregate spectrum are also pictured.}
    \label{fig:sum_change_output}
\end{figure}

As hinted at in the prior example, within the summation result, the user can also specify calculation results based on the properties of the isotopes.  
The example script {\texttt{examples/TimeEvolvingReactor.py}} is provided in CONFLUX to show that, within a reactor neutrino summation calculation, one can obtain the individual or aggregate spectra of a specific subgroup of isotopes based on their end-point energy, fission fraction, or uncertainty contribution.  
The outputs of this example, meant to show individual reactor neutrino flux contributions from fission products with Q-values above 10~MeV, are shown in figure~\ref{fig:sum_change_output}.  
The example showed CONFLUX can output the neutrino spectra of isotopes that contribute in a highlighted energy range, or that have a large uncertainty impact to the whole isotopic reactor neutrino flux calculation.  
Such selected output modes can benefit phenomenological studies to understand data-model discrepancies in reactor neutrino flux.
    
\subsection{Beta conversion - ILL measurement conversion}
\begin{figure}[H]
    \centering
    \includegraphics[width=0.49\textwidth]{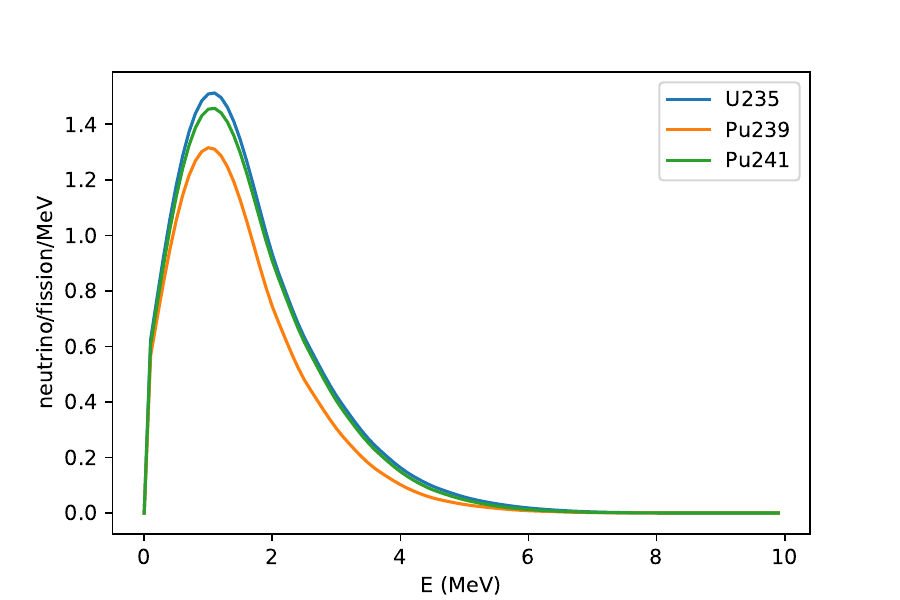}
    \includegraphics[width=0.49\textwidth]{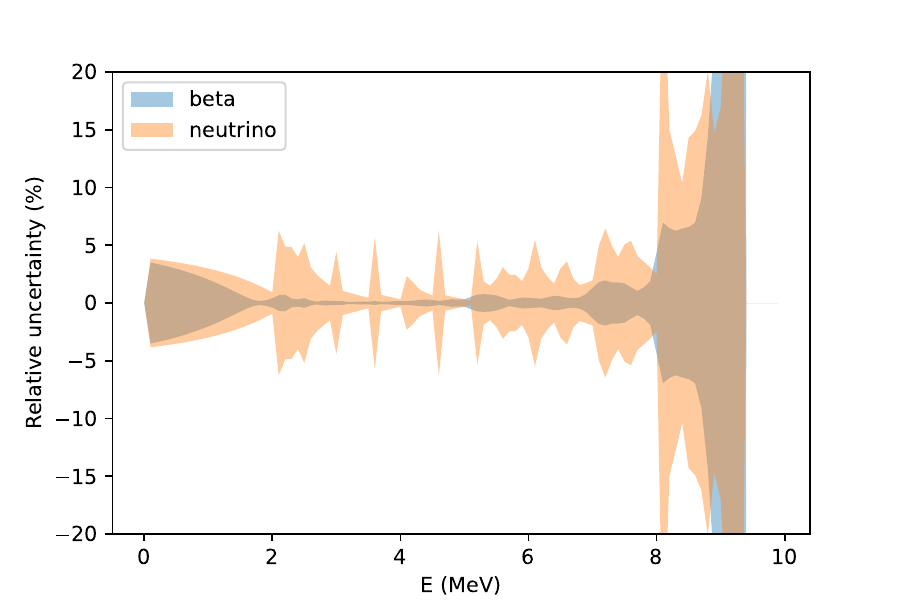}
    \caption{Left: the ILL-measured \ce{^235U}, \ce{^239Pu}, and \ce{^241Pu}~aggregate fission beta spectrum, along with fitted individual and aggregate beta branches and the aggregate converted neutrino spectrum.  Right: uncertainties associated with the aggregate fitted beta and converted neutrino spectra.}
    \label{fig:conversion_235}
\end{figure}

The example {\texttt{BetaConversionExample.py}} in the CONFLUX package shows the user how the conversion calculation is processed with the default historical ILL measurement data.
The example lets user calculate the neutrino spectra by fitting the beta spectra of \ce{^235U}, \ce{^239Pu}, and \ce{^241Pu}, with results shown in figure~\ref{fig:conversion_235}.  
Specifically plotted in the left hand panel are the fitted aggregate beta spectra, individual fitted virtual beta branch spectra, and the aggregate converted neutrino spectrum.  
In this example, the sizes of spectrum slice was changed from 0.5 MeV to 0.25 MeV to indicate the fitting results with different numbers of virtual beta spectra.
Uncertainties calculated through the MC sampling process are also provided, and are shown in the right plot of figure~\ref{fig:conversion_235}.
If a new beta spectrum data was measured from fissile isotopes or a group of beta-decays, the conversion fitting can follow the exact procedure as in this example.  

\subsection{Hybrid calculation}
\begin{figure}[H]
    \centering
    \includegraphics[width=0.6\textwidth]{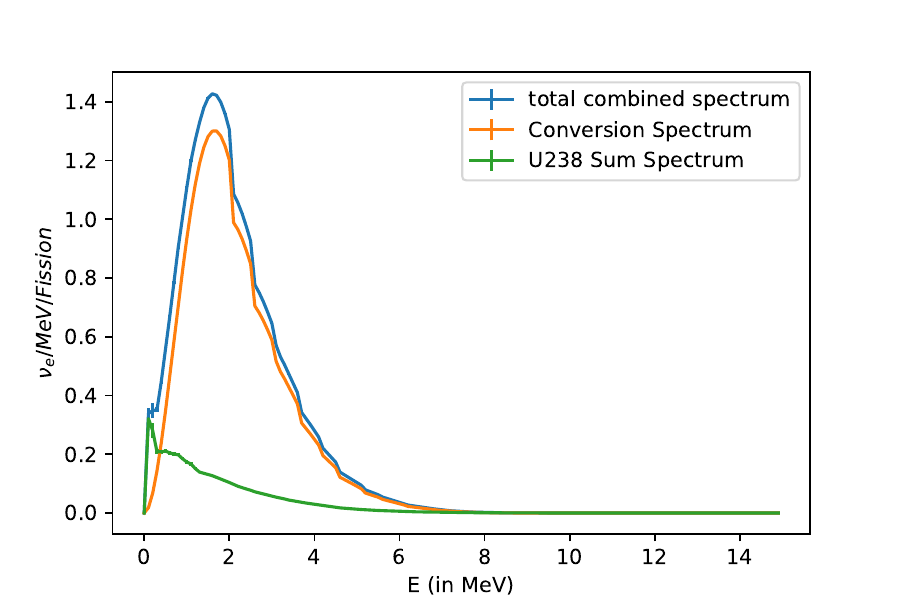}
    \caption{Adding converted neutrino spectra from\ce{^235U}, \ce{^239Pu}, and \ce{^241Pu}~with the summation-calculated \ce{^238U}~neutrino spectrum to yield a neutrino spectrum model for a LEU-like reactor.}
    \label{fig:hybrid_dyb}
\end{figure}

Combining the results of the summation mode and the conversion mode is a common approach to predict reactor neutrino fluxes.  
This method is often mainly data-driven with the conversion approach, with unmeasured spectrum composition theoretically calculated via the summation method.  
In the example {\texttt{examples/HybridReactor.py}} LEU reactor neutrino modeling is performed that contains \ce{^235U}, \ce{^239Pu}, and \ce{^241Pu}~contributions calculated using the conversion mode within CONFLUX, while \ce{^238U} was calculated using the CONFLUX summation mode, since fast neutron fission of \ce{^238U} could not be individually measured at ILL.  
The result of this hybrid calculation is shown in figure~\ref{fig:hybrid_dyb}.

\begin{figure}[H]
    \centering
    \includegraphics[width=0.6\textwidth]{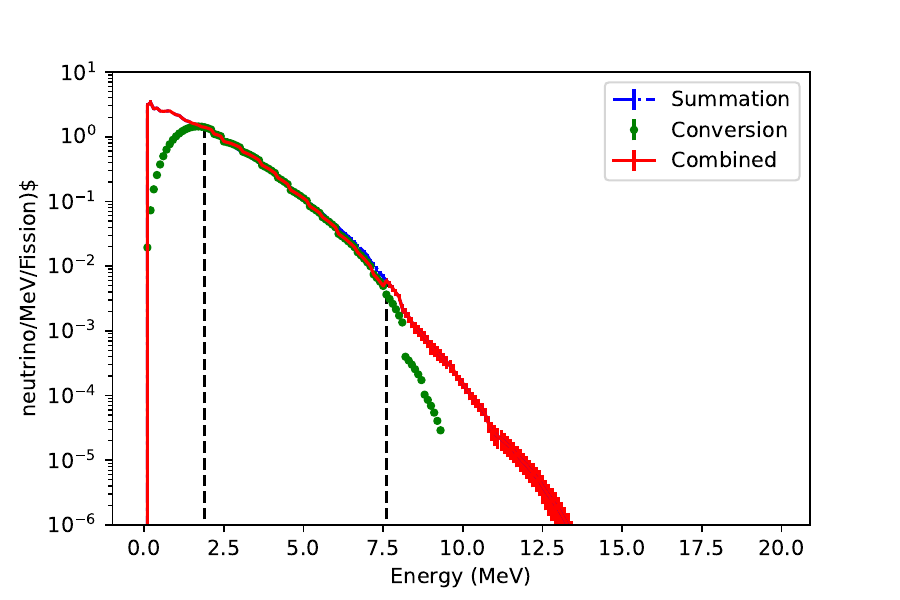}
    \caption{Adding a summed neutrino spectrum in below 1.8 MeV IBD threshold and above the 7.5 MeV range, where fitting uncertainty is high for conversion. Between the two summed range, a converted neutrino spectrum takes the place in the 1.8 MeV to 7.5 MeV. }
    \label{fig:hybrid_low_E}
\end{figure}

Another neutrino prediction mode combination application can be bound in reactor-based studies of coherent neutrino-nucleus scattering.  Since this neutrino interaction process has no energy threshold, reactor-based detectors using this channel are sensitive to the full range of produce neutrino energies -- including those outside of the range that can be predicted via the conversion approach with existing fission beta spectrum measurements.  
The CONFLUX example shown in {\texttt{example/BetaPlusSummation.py}} and pictured in figure~\ref{fig:hybrid_low_E} generates a different type of hybrid neutrino spectra spanning the full neutrino spectrum that splices together results from different modes in different energy regimes.  
Specifically, the converted neutrino spectrum is used over most of the neutrino energy range, where it provides lower reported uncertainties, while the higher-uncertainty summation-mode neutrino spectrum prediction is used below 1.8 MeV and above 8 MeV, where the conversion approach is unable to deliver a meaningful prediction.  

\subsection{Beta/neutrino spectrum with non-zero neutrino mass}
\begin{figure}[H]
    \centering
    \includegraphics[width=0.5\linewidth]{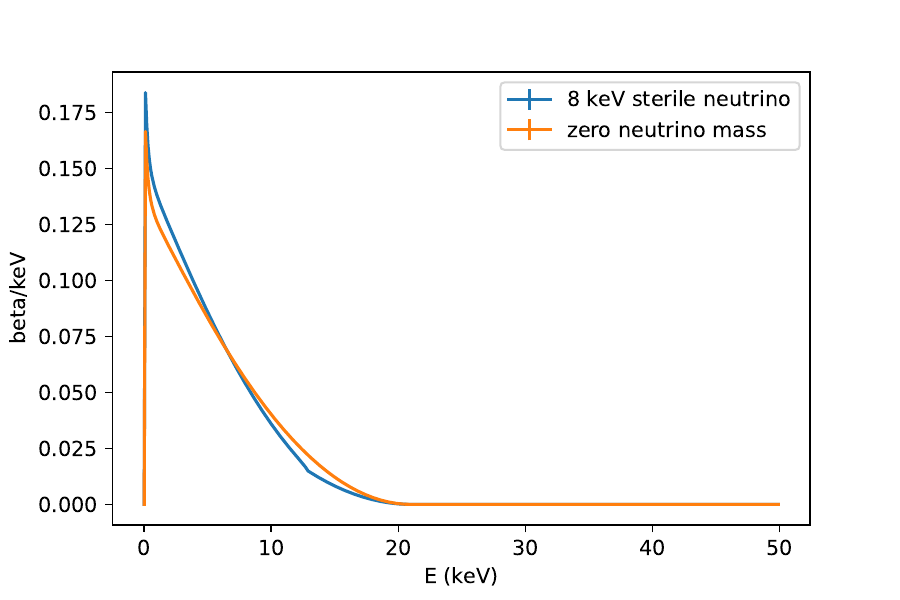}
    \caption{Beta decay spectrum of \puone. The comparison is made between the massless neutrino case and an 8~keV sterile neutrino with massively exaggerated 50\% mixing.}
    \label{fig:numass}
\end{figure}

Studying unknown properties such as neutrino mass or oscillations has been a key topic for experimental neutrino measurements.  
Famous spectral shape measurements for neutrino physics include the sterile neutrino mixing measurement of the KATRIN tritium endpoint experiment~\cite{PhysRevLett.126.091803}, or the planned JUNO measurement of the oscillated reactor neutrino energy spectrum shape to determine the neutrino mass hierarchy~\cite{cabrera_synergies_2022}.  
CONFLUX provides a way to modify the phase space calculation with a non-zero neutrino mass and flavor mixing caused by the presence of a single mass difference between two neutrino mass eigenstates.  
Currently, CONFLUX has not implemented methods for handling a more complicated neutrino mass mixing model that contains multiple neutrino mass differences, their corresponding PMNS matrix elements, or other more complex standard or non-standard flavor transformation effects.
The simplified calculation that can only include a single neutrino mass or a single mass splitting is sufficient for application of CONFLUX to a beta-spectrum-based neutrino mass measurement use case, or to a use case of a reactor-based short-baseline neutrino oscillation experiment. 
CONFLUX provides one example at {\texttt{examples/pu241\char`_numass.py}} to calculate the distortion or difference at the spectrum end-point of the \ce{^241Pu}~beta-decay spectrum due to non-zero neutrino masses, as displayed in figure~\ref{fig:numass}.  

\subsection{Time dependent neutrino flux of independent fission products}
\begin{figure}[H]
    \centering
    \includegraphics[width=0.5\linewidth]{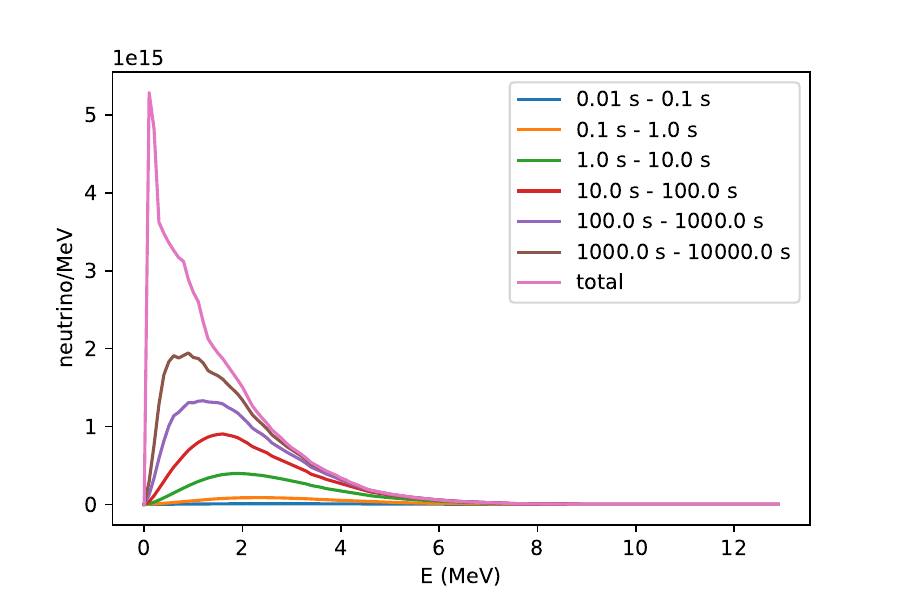}
    \caption{Cumulative neutrino spectrum in different windows of time after a \ce{^235U}~ triggered by 0.4 MeV neutron. Neutrinos from the higher order decay products are ignored from this calculation.}
    \label{fig:nuexplosion}
\end{figure}

Neutrinos generated from pulse fission can is of interests for the CEvNS detection from a pulse fission reactor~\cite{ANG2021165342}.
The modelling of neutrinos from burst nuclear fissions, such as nuclear explosions, is important to the potential application of neutrino detection for nuclear security and safeguard~\cite{annurev-nucl-102122-023751}. 
Fission reactions from these sources happens in the ms scale, bring the challenge that fission products do not reach equilibrium. 
Calculating neutrino flux of these sources requires independent FPYs from the nuclear database such as ENDF and JEFF, correlation of the fission products, as well as the beta/neutrino spectrum.
The example \textbf{examples/fission\char`_burst.py} shown in figure~\ref{fig:nuexplosion} indicate the calculation of neutrino produced in incrementing time windows after a imaginary burst of \ce{^235U}~fission. 
The neutrino spectra in each period are the summation of neutrino spectra fission products, weighted based on the time passed after the fission and  the fission products' half-lives. 

Currently, CONFLUX does not calculate the secondary and higher order decay products from the decay chains of the independent fission products, due to some decay chains contain more types of decay other than beta decay. An update of CONFLUX will tackle this issue.

\section{Conclusion}
\label{sec:conclusion}
CONFLUX was created with the goal to provide a framework for the neutrino physics field to quickly model the neutrino flux from a reactor or from beta-decays for a variety of experimental or theoretical studies.
The software uses a Python framework that contains methods to calculate reactor neutrino fluxes based on the BSG beta-decay theory engine.  
Nuclear databases are parsed into simplified databases using the xml or csv format and are loaded so neutrino spectra can be summed with respect to known fission yields and nuclear structure information. 
The calculation methods to convert an aggregate fission beta spectrum measurement into a predicted neutrino spectrum are also provided to enable users to easily calculate a data-driven neutrino flux prediction.  
Uncertainties accompanying the various prediction methods, including uncertainty covariance among beta branches and fission fragments, can also be delivered by CONFLUX.
Many input attributes determining a final predicted neutrino spectrum, from chosen input nuclear databases to the fundamental beta-decay theory assumptions, can be customized in CONFLUX to assist users in studying the potential modifications to reactor neutrino flux predictions.
CONFLUX can be used as a tool to quickly calculate the neutrino flux from various reactors to aid users in experimental design or to generate reference models for a fundamental or applied reactor neutrino measurement or a neutrino-physics related nuclear-physics measurement.
CONFLUX will be continuously improved and updated with new functions, including a future neutrino data-driven calculation mode, calculations of $\beta+$ decay spectra, and more complete modeling of the complicated decay chains of fission products when generating neutrino predictions from independent fission product yield inputs.  

\section*{Acknowledgements}

This work was supported by the Lawrence Livermore National Laboratory LDRD Program under Project No. 20-SI-005, the U.S. Department of Energy Office of Science, Office of High Energy Physics under Award No. DE-SC0020262 to Virginia Polytechnic Institute and State University and under Work Proposal Number SCW1504 to Lawrence Livermore National Laboratory, and by the U.S. Department of Energy Office of Defense Nuclear Nonproliferation Research and Development.  This work was supported by the Consortium for Monitoring, Technology, and Verification under DOE-NNSA award number DE-NA0003920. The authors thank Daniel Nestares from the University of California, Merced, for his work to test the software. The authors thank Mitchel Crockett from the University of Tennessee, Knoxville, for his reactor simulation output to aid CONFLUX on understanding the reactor simulation input. At last, The authors thank Eric F. Matthew from University of California, Berkeley for providing a reference fission product covariance dataset. This work was performed under the auspices of the U.S. Department of Energy by Lawrence Livermore National Laboratory under Contract DE-AC52-07NA27344. LLNL-JRNL-872396.





\bibliographystyle{elsarticle-num} 

\bibliography{main.bib}








\end{document}